\documentclass[11pt,preprint]{aastex}

\usepackage{color}
\usepackage{natbib}
\usepackage{units}
\usepackage{amsmath}
\usepackage{grffile}     
\tightenlines
\slugcomment{}

\newcommand \fCgra      {f_{\rm C,gra}}
\newcommand \abs        {{\rm abs}}

\newcommand \Angstrom   {\,{\rm \AA}}

\newcommand \bahat      {\hat{\bf a}}
\newcommand \bB         {{\bf B}}

\newcommand \bchat      {\hat{\bf c}}

\newcommand \bE         {{\bf E}}
\newcommand \behat      {\hat{\bf e}}

\newcommand \bkhat      {\hat{\bf k}}

\newcommand \bxhat      {\hat{\bf x}}
\newcommand \byhat      {\hat{\bf y}}
\newcommand \bzhat      {\hat{\bf z}}

\newcommand \beq        {\begin{equation}}
\newcommand \beqa	{\begin{eqnarray}}

\newcommand \cm         {\,{\rm cm}}

\newcommand \eeq	{\end{equation}}
\newcommand \eeqa	{\end{eqnarray}}

\newcommand \ext        {{\rm ext}}
\newcommand \eV 	{\,{\rm eV}}
\newcommand \gm         {\,{\rm g}}

\newcommand \gtsim	{\gtrsim}		 

\newcommand \keV	{\,{\rm keV}}
\newcommand \khat       {\hat{\bf k}}

\newcommand \K  	{\,{\rm K}}

\newcommand \ltsim	{\lesssim}		 

\newcommand \meV        {{\,\rm meV}}

\renewcommand \mho      {\,{\rm mho}}

\newcommand \nm         {\,{\rm nm}}     

\newcommand \NH         {N_{\rm H}}

\newcommand \s	        {\,{\rm s}}
\newcommand \sca        {{\rm sca}}

\newcommand \xtimes     {{\!\,\times\!\,}}

\newcommand \aeff       {a_{\rm eff}}

\newcommand \pol        {{\rm pol}}

\newcommand \onethirdtwothird {\nicefrac{1}{3}\,-\,\nicefrac{2}{3}}

\newcommand \website    {http://www.astro.princeton.edu/$\sim$draine/dust/D16graphite/D16graphite.html}
\newcommand \websiteb   {http://arks.princeton.edu/ark:/88435/dsp01nc580q118}

\newcommand{\oldtext}[1]{}
\newcommand{\newtext}[1]{{#1}}

\countdef\decade=200
\advance\decade by \year
\countdef\hours=201
\advance\hours by \time
\divide\hours by 60
\countdef\mins=202
\advance\mins by \hours
\multiply\mins by 60
\multiply\hours by 100
\countdef\miltime=203
\advance\miltime by \hours
\advance\miltime by \time
\advance\miltime by -\mins

\begin{document}

\title{%
        \vspace*{-3.0em}
        {\normalsize\rm To appear in {\it The Astrophysical Journal}}\\ 
                \vspace*{1.0em}
        Graphite Revisited
	}

\author{
B.\ T.\ Draine
\\
Princeton University Observatory, Peyton Hall, Princeton, NJ 08544-1001, USA;
draine@astro.princeton.edu}

\begin{abstract}
Laboratory measurements are used to constrain the dielectric
tensor for graphite, from microwave to X-ray frequencies.
The dielectric tensor is strongly anisotropic even at X-ray energies.
The discrete dipole approximation is employed for accurate
calculations of absorption and scattering by single-crystal graphite spheres
and spheroids.
For randomly-oriented
single-crystal grains, the so-called \onethirdtwothird\ approximation 
for calculating absorption and scattering cross sections is
exact in the limit $a/\lambda\rightarrow 0$, provides 
better than $\sim$10\% accuracy in the optical and UV
even when $a/\lambda$ is not small,
but becomes increasingly inaccurate at infrared wavelengths,
with errors as large as $\sim$40\% at $\lambda=10\micron$.
For turbostratic graphite grains, the Bruggeman and Maxwell Garnett treatments
yield similar cross sections in the optical and ultraviolet, but diverge
in the infrared, with predicted cross sections differing by over an
order of magnitude in the far-infrared.  It is argued that the Maxwell Garnett
estimate is likely to be more realistic, and is recommended.
The out-of-plane lattice resonance of graphite
near $11.5\micron$ may be observable in absorption
with the MIRI spectrograph on JWST. 
Aligned graphite grains, if present in the ISM, 
could produce polarized X-ray absorption and
polarized X-ray scattering near the carbon K edge.
\end{abstract}
\keywords{solid state: refractory ---
          ISM: dust, extinction ---
          infrared: ISM ---
          submillimeter: ISM ---
          ultraviolet: ISM ---
          X-rays: ISM}

\section{Introduction
         \label{sec:intro}}

First proposed as an interstellar grain material by
\citet{Cayrel+Schatzman_1954} and
\citet{Hoyle+Wickramasinghe_1962}, 
the graphite hypothesis received support with the
discovery by \citet{Stecher_1965}
of a strong extinction ``bump'' at $2175\Angstrom$, 
consistent with the absorption calculated for small graphite spheres
\citep{Stecher+Donn_1965}.
A number of other carbonaceous materials have also been proposed
as important constituents of the interstellar dust population,
including
polycyclic aromatic hydrocarbons \citep{Leger+Puget_1984,
Allamandola+Tielens+Barker_1985},
hydrogenated amorphous carbon \citep{Duley+Jones+Williams_1989},
amorphous carbon \citep{Duley+Jones+Taylor+Williams_1993},
fullerenes \citep{Webster_1992,Foing+Ehrenfreund_1994},
and
diamond \citep{Hill+Jones+DHendecourt_1998,Jones+DHendecourt_2004}.

The total abundance of carbon in the interstellar medium (ISM) 
has been estimated to be ${\rm C/H}=\newtext{339\pm41}$\,ppm
\citep{Asplund+Grevesse+Sauval+Scott_2009},
although other estimates range from 
$214\pm20$\,ppm \citep{Nieva+Przybilla_2012}
to $>464\pm57$\,ppm \citep{Parvathi+Sofia+Murthy+Babu_2012}.
In the diffuse ISM, $\sim 40-70\%$ of the carbon is in
C$^+$, C$^0$, or small molecules such as CO, CN, CH, and CH$^+$.
The remainder ($\sim 30-60\%$) of the carbon
is in grains, extending down to nanoparticles
containing as few as $\sim$$20$ C atoms.
However, the physical forms in which this carbon is present
remain uncertain.

The presence of ${\rm C}_{60}^+$ in the diffuse ISM was recently
confirmed 
\citep{Campbell+Holz+Gerlich+Maier_2015,Walker+Bohlender+Maier+Campbell_2015},
but accounts for only $\sim$0.05\%
of the carbon on the sightlines studied; the entire fullerene family
(C$_{60}$, C$_{60}^+$, C$_{70}$,
C$_{70}^+$, ...) probably accounts for $\ltsim 0.2\%$ of the interstellar
carbon.

The strong interstellar extinction feature at $2175\Angstrom$ continues to
point to $sp^2$-bonded carbon 
in aromatic rings (as in graphite).
With the oscillator strength per carbon estimated to be $f\approx 0.16$
\citep{Draine_1989a}, $\sim$20\% of the interstellar carbon is required
to produce the observed $2175\Angstrom$ feature.

The strong mid-infrared emission features at 3.3, 6.2, 7.7, 8.6, 11.3, and
12.7$\micron$ appear to be radiated by 
polycylic aromatic hydrocarbon (PAH) nanoparticles, in which the
C atoms are organized in hexagonal (aromatic) rings, just as in graphite.
While a portion of the carbon-carbon bonds in the mid-infrared emitters
could be ``aliphatic'' (such as open-chain hydrocarbons), 
the emission spectra appear to show that a majority of the carbon-carbon bonds 
are ``aromatic''
\citep{Li+Draine_2012,Yang+Glaser+Li+Zhong_2013}
\citep[but see also][]{Kwok+Zhang_2011,Kwok+Zhang_2013,
Jones+Fanciullo+Kohler+etal_2013}.
Estimates for the fraction of the carbon contained in
PAHs range from $\sim$7\%
\citep{Tielens_2008} to $\sim$20\% \citep{Li+Draine_2001a,Draine+Li_2007}.
It now seems likely that much -- perhaps most -- of the $2175\Angstrom$
feature
is produced by the $sp^2$-bonded carbon in
the nanoparticles responsible for the 3.3--12.7$\micron$ emission
features 
\citep{Leger+Verstraete+dHendecourt+etal_1989,Joblin+Leger+Martin_1992}. 

It remains unclear what form the remainder of the
carbon is in.
Graphite and diamond are the two crystalline states of pure carbon;
graphite is the thermodynamically favored form at low pressures.
However, many forms of ``disordered'' carbon materials exist, including
``glassy'' carbons and hydrogenated amorphous carbons
\citep{Robertson_1986}.

An absorption feature at $3.4\micron$ is identified as
the C-H stretching mode in aliphatic (chainlike) 
hydrocarbons, but 
the fraction of the carbon that must be aliphatic to account for 
the observed feature is uncertain.
Based on the 3.4$\micron$ feature,
\citet{Pendleton+Allamandola_2002} estimated that interstellar carbonaceous
material was $\sim$85\% aromatic and $\sim$15\% aliphatic
\citep[but see][]{Dartois+MunozCaro+Deboffle+dHendecourt_2004}.
Papoular argues that the carbonaceous material in the ISM
resembles
coals \citep{Papoular+Breton+Gensterblum+etal_1993}
or kerogens \citep{Papoular_2001} -- 
disordered macromolecular materials with much
of the carbon in aromatic form, but with a significant
fraction of the carbon in nonaromatic forms, containing substantial
amounts of hydrogen, and a small amount of oxygen.
\citet{
Jones_2012d,
Jones_2012e,
Jones_2012a,
Jones_2012b,
Jones_2012c}
considers the carbon in interstellar grains to be in a range of
forms: the outer layers (``mantles'') are highly aromatic, 
the result
of prolonged UV irradiation, but much of the carbon is in
grain interiors, in the form of ``hydrogenated amorphous carbon'' (aC:H),
a nonconducting material with a significant bandgap.

The evolution of interstellar dust is complex and as-yet poorly
understood.
The objective of the present study is not to argue for or against
graphite as a constituent of interstellar grains, but rather to
provide an up-to-date discussion of the optical properties of graphite
for use in modeling graphite particles that may be
present in the ISM
or in some stellar outflows.

The plan of the paper is as follows.
In \S\ref{sec:epsilon}-\ref{sec:Kshell}
the laboratory data are reviewed, and a dielectric
tensor is obtained that is generally consistent with published laboratory
data (which are themselves not all mutually consistent),
including measurements of polarization-dependent X-ray absorption near the
carbon K edge.
Because graphite is an anisotropic material, calculating absorption
and scattering by graphite grains presents technical challenges.
Techniques for calculating absorption and scattering by
single-crystal spheres and spheroids are discussed in
\S\ref{sec:cross sections}; accurate cross sections obtained with
the discrete dipole approximation are used to test the so-called
\onethirdtwothird\ approximation.
Effective medium theory approaches for modeling turbostratic
graphite grains are discussed and compared in \S\ref{sec:EMT}.
In \S\ref{sec:emt cross sections} we present selected results for extinction
and polarization cross sections for turbostratic graphite
spheres and spheroids, as well
as Planck-averaged cross sections for absorption and radiation
pressure.
Observability of the
out-of-plane $11.5\micron$ lattice resonance is discussed in
\S\ref{sec:lattice resonance}.
In \S\ref{sec:X-ray} it is shown that
graphite grains in the ISM, if aligned, will polarize the
$280-330\eV$ radiation reaching us from X-ray sources; 
the scattered X-ray ``halo''
will also be polarized.
The results are discussed in \S\ref{sec:discussion} and summarized
in \S\ref{sec:summary}.

\section{\label{sec:epsilon}
         Dielectric Tensor for Graphite}

In graphite, with density $\rho=2.26\gm\cm^{-3}$,
the carbon atoms are organized in 2-dimensional
graphene sheets.
Within each graphene sheet, the atoms are organized in a hexagonal
lattice, with nearest-neighbor spacing $1.42\Angstrom$.
The bonding between sheets is weak.
The sheets are
stacked according to several possible
stacking schemes,
with interlayer spacing $d=3.35\Angstrom$.

Carbon has 4 electrons in the $n=2$ shell.
In graphene or graphite, three electrons per atom
combine in $\sigma$ orbitals forming coplanar carbon-carbon bonds
(so-called ``$sp^2$ bonding'');
the remaining
valence electron is in a $\pi$ orbital, extending above and
below the plane.  This higher-energy $\pi$ orbital is responsible
for the electrical conductivity.  Because the top of the $\pi$
valence band overlaps slightly with the bottom of the
conduction band, graphite is a ``semimetal'', with
modest electrical conductivity even at low temperatures.

Graphite's structure
makes its electro-optical properties extremely anisotropic.
Graphite is a uniaxial crystal; the ``{\it c}-axis'' $\bchat$ is normal 
to the basal plane (i.e., the graphene layers).
The dielectric tensor has two components: 
$\epsilon_\parallel(\omega)$
describing the response to electric fields $\bE\parallel \bchat$,
and $\epsilon_\perp(\omega)$
for the response when $\bE\perp \bchat$.

While large crystals of natural graphite have been used for
some laboratory studies 
\citep[e.g.,][]{Soule_1958, Greenaway+Harbeke+Bassani+Tosatti_1969},
most work employs
the synthetic material
known as ``highly oriented pyrolitic graphite'' (HOPG).
High-quality HOPG samples consist
of graphite microcrystallites with diameters typically in the range
1--10$\micron$ (larger than typical interstellar grains)
and $\bchat$ axes aligned
to within $\sim$0.2$^\circ$ \citep{Moore_1973}.

Determination of the optical constants of graphite has proved
difficult, with different studies often obtaining
quite different results
\citep[see the review by][]{Borghesi+Guizzetti_1991}.
The optical constants for graphite are generally determined through
measurements of the reflectivity, or by 
electron energy loss spectroscopy (EELS)
on electron beams traversing the sample
\citep{Daniels+Festenberg+Raether+Zeppenfeld_1970}.
Electron emission from the sample has been used to measure
absorption at X-ray energies (see \S\ref{sec:Kshell}).

Reflectivity studies are most easily done on samples cleaved along
the basal plane, with $\bchat$ normal to the sample surface.
Normal incidence light then samples only
$\epsilon_\perp$, but at other angles the polarization-dependent
reflectivity depends on both $\epsilon_\perp$ and $\epsilon_\parallel$.
Samples can also be cut to
produce a surface containing the $\bchat$-axis, which would allow
direct measurement of $\epsilon_\parallel$ from reflectivity measurements,
but 
the resulting surfaces (even after polishing)
are generally not optical-quality, hampering reflectometry.

\subsection{Modeling Dielectric Functions}

In a Cartesian coordinate system with the $\bxhat$ and $\byhat$ axes
lying in the graphite basal plane (i.e, $\bzhat \parallel\bchat$),
the dielectric tensor is diagonal with elements
$(\epsilon_\perp,\epsilon_\perp,\epsilon_\parallel)$.
$\epsilon_\perp$ and $\epsilon_\parallel$
must each satisfy
the Kramers-Kronig relations 
\citep[see, e.g.,][]{Landau+Lifshitz+Pitaevskii_1993, Bohren_2010},
e.g., $\epsilon_1\equiv {\rm Re}(\epsilon)$ can
be obtained from $\epsilon_2\equiv {\rm Im}(\epsilon)$:
\beq \label{eq:KK}
\epsilon_1(\omega) -1 = 
\frac{2}{\pi} P \int_0^\infty 
\frac{x \epsilon_2(x)}{x^2-\omega^2} 
dx
~~~,
\eeq
where $P$ denotes the principal value.
A general approach to obtaining a Kramers-Kronig compliant 
dielectric function is to try to adjust $\epsilon_2(\omega)\geq 0$ 
to the observations,
obtaining $\epsilon_1(\omega)$ using Eq.\ (\ref{eq:KK}).
This procedure was used, for example, by \citet{Draine+Lee_1984}.

Because an analytic representation of $\epsilon(\omega)$ using a
modest number of adjustable parameters has obvious advantages, in the
present work we model the dielectric function as the sum of $N_f$
free-electron-like components, and $N_r$ damped-oscillator-like
components, plus a contribution $\delta\epsilon_{\rm K}(\omega)$ from
the K shell electrons: 
\beqa 
\label{eq:epsilon} 
\epsilon(\omega) - 1 &=& \sum_{j=1}^{N_f}
\frac{-A_j(\omega_{pj}\tau_j)^2}{(\omega\tau_j)^2 + i\omega\tau_j} +
\sum_{r=1}^{N_r}
\frac{S_r}{1-(\omega/\omega_{0r})^2-i\gamma_{r}(\omega/\omega_{0r})} +
\delta\epsilon_{\rm K}(\omega) 
\\ 
{\rm Re}(\epsilon) -1 &=&
\sum_{j=1}^{N_f} \frac{-A_j(\omega_{pj}\tau_j)^2} {1+(\omega\tau_j)^2}
+ \sum_{r=1}^{N_r} \frac{\left[1-(\omega/\omega_{0r})^2\right]S_r}
{[1-(\omega/\omega_{0r})^2]^2+\gamma_r^2(\omega/\omega_{0r})^2} + {\rm
  Re}(\delta\epsilon_{\rm K}) 
\\
{\rm Im}(\epsilon) &=& \sum_{j=1}^{N_f}
\frac{A_j(\omega_{pj}\tau_j)^2} {(\omega\tau_j)+(\omega\tau_j)^3} +
\sum_{r=1}^{N_r} \frac{S_r\gamma_r(\omega/\omega_{0r})}
    {[1-(\omega/\omega_{0r})^2]^2+\gamma_r^2(\omega/\omega_{0r})^2} +
    {\rm Im}(\delta\epsilon_{\rm K})
~~~,
\eeqa
where $A_j=\pm 1$ depending on whether the free-electron-like component
contributes a positive (i.e., physical) or negative (nonphysical)
conductivity.\footnote{Negative conductivity components are allowed
  purely to improve the overall fit to the data, but of course the
  total conductivity must be positive at all frequencies.}  
Each free-electron-like component is characterized by a plasma
frequency $\omega_{pj}$ and mean free time $\tau_j$.
Each damped oscillator component is characterized by a resonant
frequency $\omega_{0r}$, dimensionless damping parameter $\gamma_r$,
and strength $S_r$.
The K-shell contribution $\delta\epsilon_{\rm K}(\omega)$ is discussed in
\S\ref{sec:Kshell}, but can be approximated as $\delta\epsilon_{\rm K}
\approx 0.0019$ for $h\nu<50\eV$.  
Because each term in
(\ref{eq:epsilon}) satisfies the Kramers-Kronig relation
(\ref{eq:KK}), the sum does so as well.

The dielectric function satisfies various sum rules
\citep{Altarelli+Dexter+Nussenzveig+Smith_1972}, including
\beq \label{eq:neff}
Z_{\rm eff}(\omega) = \frac{1}{n_a}\frac{m_e}{2\pi^2 e^2} \int_0^\omega 
\omega^\prime \epsilon_2(\omega^\prime) d\omega^\prime
~~~,
\eeq
where $e$ and $m_e$ are the electron charge and mass,
$n_a$ is the atomic number density in the material,
and 
$Z_{\rm eff}(\omega)$ 
is the effective number of electrons per atom contributing
to absorption at frequencies $<\omega$.
We can integrate each of the model components to
obtain the total oscillator strength (i.e., number of electrons) 
associated with each component:
\beqa
f_{pj} &=& \frac{1}{n_{\rm C}} A_j
\frac{ m_e\omega_{pj}^2}{4\pi e^2}
\\
f_{r} &=& \frac{1}{n_{\rm C}} S_r
\frac{m_e\omega_{0r}^2}{4\pi e^2}
~~~.
\eeqa
The $1s^2$ electrons do not absorb at energies $\hbar\omega < 280\eV$.
Thus, if the material were isotropic
we would expect
$Z_{\rm eff}(\hbar\omega=280\eV)\approx \sum_j f_{pj} + \sum_r f_{r} \approx 4$ for both $\epsilon_\perp$ and $\epsilon_\parallel$.

In Gaussian (a.k.a.\ ``cgs'') electromagnetism, 
the electrical conductivity\footnote{$1\mho\cm^{-1}
=(2.99793)^2\xtimes10^{11}\s^{-1}$.}

\beq
\sigma(\omega)=\frac{\omega\epsilon_2(\omega)}{4\pi}
~~~.
\eeq
The low-frequency conductivity 
$\sigma_{\rm dc}\equiv\sigma(\omega\rightarrow 0)$
is due to the free-electron-like components:
\beqa
\sigma_{\rm dc} &=& \sum_{j=1}^{N_f} \sigma_{{\rm dc},j}
\quad , \quad
\sigma_{{\rm dc},j} \equiv \frac{A_j\omega_{pj}^2}{4\pi} \tau_j
~~~.
\eeqa
\subsection{Small-particle effects}

Laboratory studies of graphite employ macroscopic samples, but
the properties of nanoparticles differ from bulk material.

\subsubsection{Surface States}

The electronic wavefunctions near the surface will differ from
the wavefunctions within the bulk material, leading to
changes in $\epsilon_\perp$ and $\epsilon_\parallel$ at all energies. 
These surface states affect laboratory measurements
of reflectivity, but reflectometry probes the dielectric function
throughout a layer of thickness $\sim\lambda/|m|$, where
$m=\sqrt{\epsilon}$ is the complex refractive index,
extending well beyond the surface monolayer.
Little appears to be known about how $\epsilon_\perp$
or $\epsilon_\parallel$ may behave close to the surface.

\subsubsection{Changes in Band Gap}
Bulk graphite is a semimetal: 
the $\pi$ electron valence band slightly overlaps the 
(nominally empty) conduction
band, resulting in nonzero electrical conductivity and semimetallic behavior
for $\bE\perp\bchat$, even at low temperatures.
In contrast,
single-layer graphene is a semiconductor with zero bandgap $E_g=0$,
but two graphene layers are sufficient to have band overlap ($E_g<0$)
with behavior approaching that of graphite for 10 or more layers
\citep{Partoens+Peeters_2006}.
The interlayer spacing is $\sim0.335\nm$; 
thus a crystal thickness $\gtsim 3\nm$ (perpendicular to the basal plane)
appears to be sufficient to
approach the behavior of bulk graphite.

While an infinite sheet of single-layer graphene has $E_g=0$, finite-width
single-layer strips have $E_g>0$, with
\beq
E_g \approx \frac{10\meV}{(W/18\nm -1)}
\eeq
for strip width $W\gtsim 20\nm$ \citep{Han+Ozyilmaz+Zhang+Kim_2007}.
This would suggest that graphite nanoparticles with diameter $D\ltsim20\nm$
might be characterized by band gap $E_g\gtsim 50\meV$, with
absorption suppressed for $h\nu\ltsim 50\meV$, or $\lambda \gtsim 25\micron$.

Neutral PAHs with $\ltsim 10^3$ C atoms have
bandgap $E_g \approx 5.8\eV/\sqrt{M}$, where $M$ is the number
of aromatic rings \citep{Salama+Bakes+Allamandola+Tielens_1996},
and we would expect similar band structure if the H atoms were removed from
the perimeter of the PAH, leaving a fragment of graphene.
A graphene fragment of diameter $D$ would have $M\approx 15(D/\nm)^2$ rings; 
thus we might expect $E_g \approx 1.5\eV/(D/\nm)$, or $E_g\approx75\meV$ for
$D\approx20\nm$, in rough agreement with the bandgap measured
by \citet{Han+Ozyilmaz+Zhang+Kim_2007} for single-layer graphene strips.
Thus very small neutral {\it single-layer} graphene particles 
would be expected to 
have a significant band gap, 
resulting in suppressed opacity at long wavelengths.

However, \citet{Mennella+Brucato+Colangeli+etal_1998} found that
$D\approx10\nm$ amorphous carbon {\it spheres} 
absorb well at $\lambda \gtsim 1000\micron$, even at $T=25\K$ ($kT=2\meV$),
implying a band gap $E_g\ltsim 2\meV$.
The opacities measured by \citet{Mennella+Brucato+Colangeli+etal_1998}
indicate that small-particle effects do not suppress the low-frequency
absorption by amorphous carbon nanoparticles for sizes down
to $D\approx10\nm$ -- evidently the {\it stacking} of the graphene
layers lowers the bandgap sufficiently to permit absorption for
even $h\nu\ltsim1\meV$, even though one would expected {\it single-layer}
graphene with $D\approx10\nm$ to be unable to absorb for $h\nu\ltsim 50\meV$.

At this point it remains unclear how the electronic
energy levels of graphite actually change with reduced particle size,
but there is no evidence of a bandgap $E_g>0$ even for particles as small
as $D=10\nm$.

\subsubsection{Surface Scattering}

In small particles, 
the mean free time $\tau$ for the ``free'' electrons
will be reduced by scattering off the surface, leading to
changes in the dielectric function at low frequencies,
where the ``free electron'' component dominates.
Following previous work
\citep{Kreibig_1974,Hecht_1981,Draine+Lee_1984},
we approximate this effect for a grain of radius $a$ by setting
\beq
\tau_j^{-1} = \tau_{{\rm bulk},j}^{-1} + \frac{v_{\rm F}}{a}
~~~,
\eeq
where $\tau_{{\rm bulk},j}$ are given in Tables 
\ref{tab:parameters for Eperpc} and
\ref{tab:parameters for Eparac},
and $v_{\rm F}=(2E_{\rm F}/m_*)^{1/2}$ is the Fermi velocity for
Fermi energy $E_{\rm F}$ and effective mass $m_*$.
For pyrolitic graphite
$E_{\rm F}=0.022\eV$;
electrons and holes have effective masses $m_{*,e}=0.039m_e$
and $m_{*,h}=0.057m_e$
\citep{Williamson+Foner+Dresselhaus_1965}.
Thus we take $v_{\rm F}\approx 4\times10^7\cm\s^{-1}$.

\section{\label{sec:FIR_to_EUV}
         FIR to EUV}
\subsection{$\bE\perp \bchat$}

\begin{table}[t]
\begin{center}
\caption{\label{tab:data}Selected determinations of $\epsilon_\perp$ or
$\epsilon_\parallel$ for graphite.}
{\footnotesize
\begin{tabular}{c c c l}
\hline
$h\nu\,(\eV)$ & case & method & reference \\
\hline
1--26 & $\bE\perp c$ & reflectivity & 
   \citet{Taft+Philipp_1965}\\
80--700 & $\bE\perp c$, $\bE\parallel c$ & absorption &
   \citet{Fomichev+Zhukova_1968} in 
   \citet{Hagemann+Gudat+Kunz_1974,Hagemann+Gudat+Kunz_1975}\\
2--5  & $\bE\perp c$, $\bE\parallel c$& reflectivity &
   \citet{Greenaway+Harbeke+Bassani+Tosatti_1969}\\
3--35 & $\bE\perp c$ & EELS &
\citet{Tosatti+Bassani_1970}\\
6--30 & $\bE\parallel c$ & EELS &
\citet{Tosatti+Bassani_1970}\\
3--40 & $\bE\perp c$, $\bE\parallel c$ & reflectivity &
   \citet{Klucker+Skibowski+Steinmann_1974}\\
1--40 & $\bE\perp c$, $\bE\parallel c$ & EELS &
   \citet{Venghaus_1975}\\
0.06--0.50 & $\bE\perp c$, $\bE\parallel c$ & reflectivity &
   \citet{Nemanich+Lucovsky+Solin_1977}\\
0.001--1.0 & $\bE\perp c$ & reflectivity &
   \citet{Philipp_1977}\\
0.012--0.50 & $\bE\perp c$, $\bE\parallel c$ & reflectivity &
   \citet{Venghaus_1977}\\
275--345 & $\bE\perp c$, $\bE\parallel c$ & absorption &
   \citet{Rosenberg+Love+Rehn_1986}\\
1.65--3.06 &$\bE\perp c$, $\bE\parallel c$ & reflectivity &
   \citet{Jellison+Hunn+Lee_2007}\\
0.006--4 & $\bE\perp c$ & reflectivity &
\citet{Kuzmenko+vanHeumen+Carbone+vanderMarel_2008,Papoular+Papoular_2014}\\
\hline
\end{tabular}
}
\end{center}
\end{table}

Graphite conducts relatively well in the basal plane.
High quality natural crystals have
measured d.c. conductivities  $2.5\times10^4\mho\cm^{-1}$ at
$T=300\K$, rising to $\sim$$2\times10^5\mho\cm^{-1}$
at $T=20\K$ \citep{Soule_1958}.

\begin{figure}[ht]
\begin{center}
\includegraphics[angle=0,width=8.0cm,
                 clip=true,trim=0.5cm 0.5cm 0.5cm 0.5cm]
{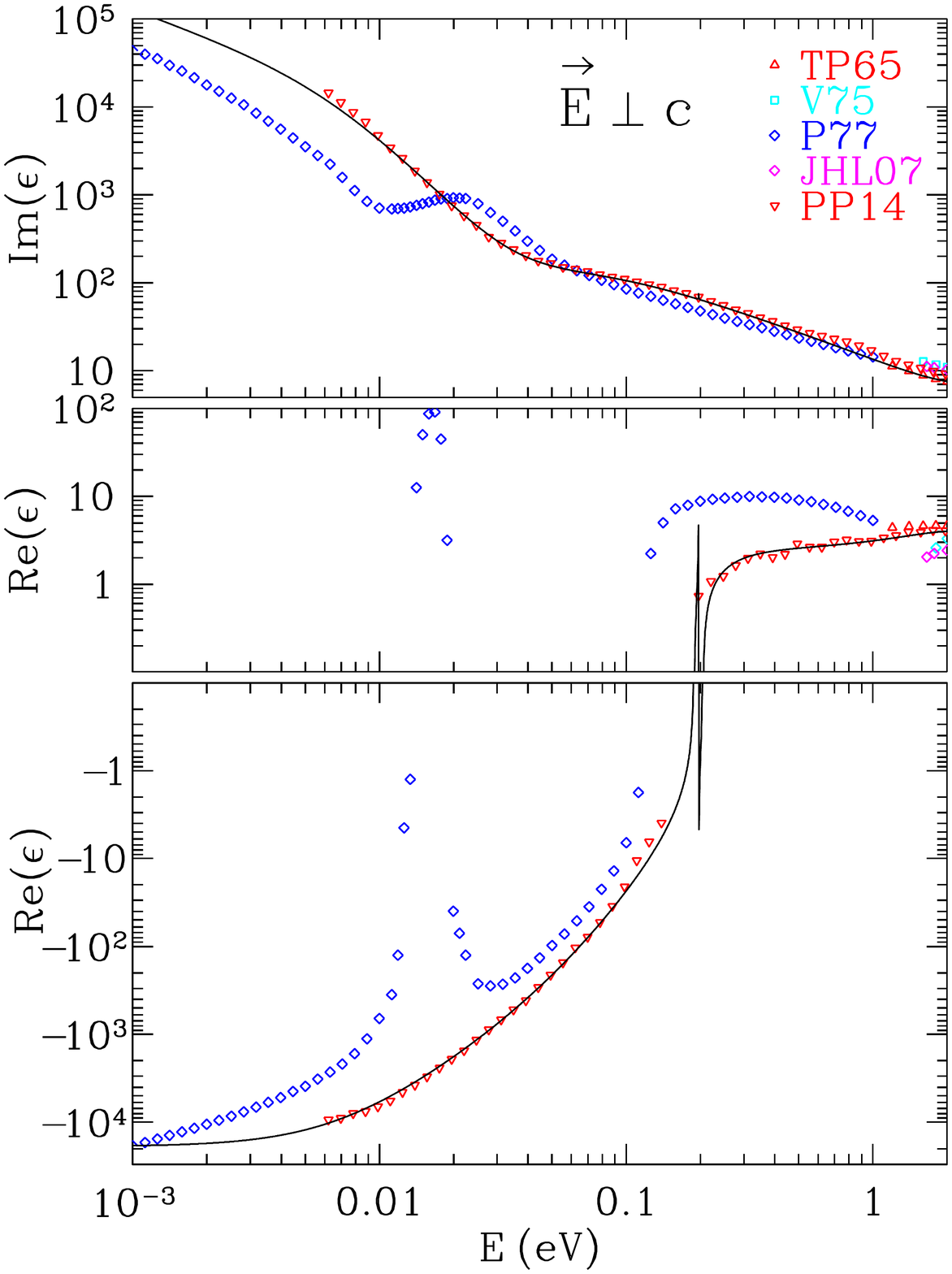}
\includegraphics[angle=0,width=8.0cm,
                 clip=true,trim=0.5cm 0.5cm 0.5cm 0.5cm]
{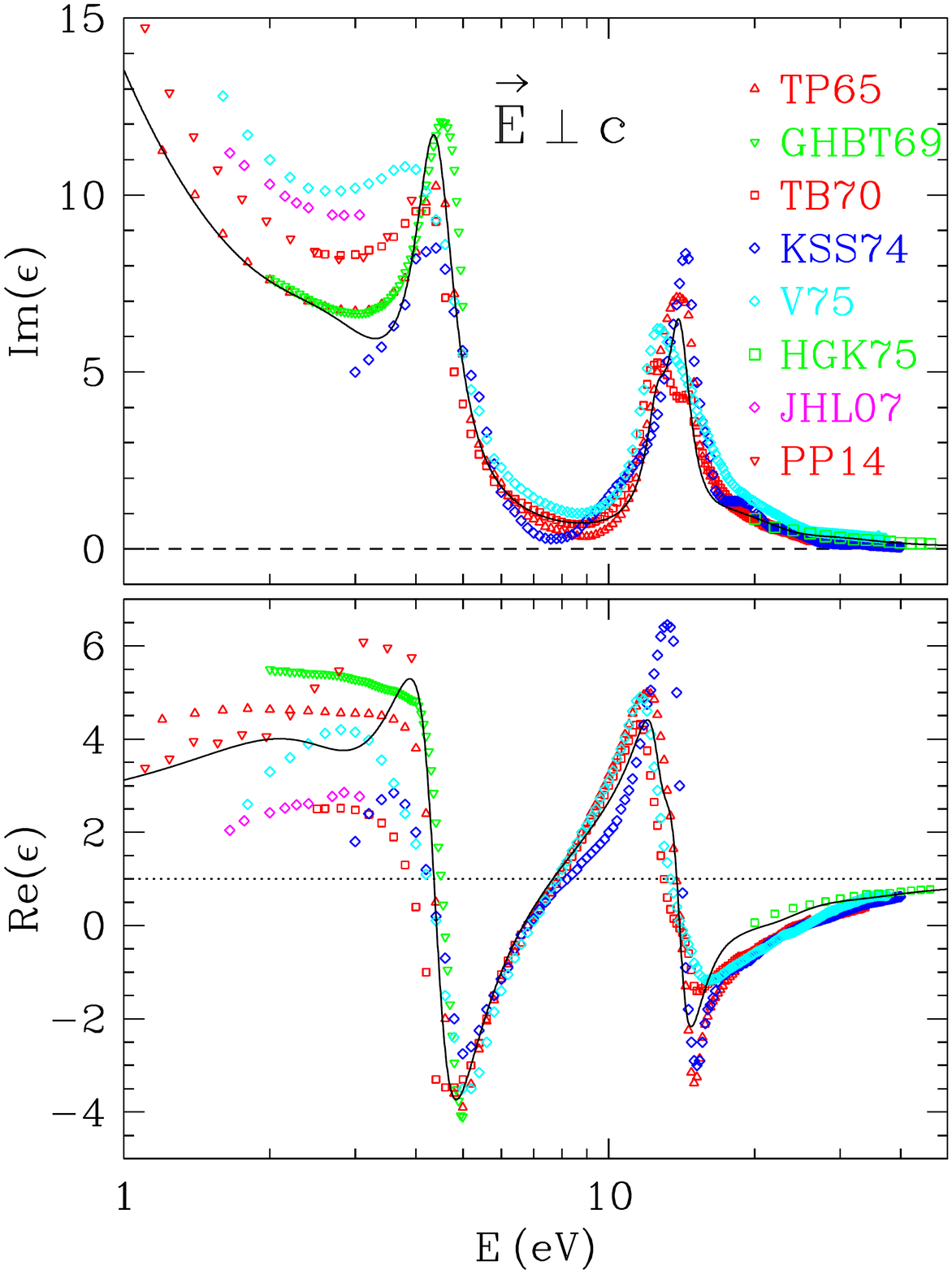}
\caption{\label{fig:eps Eperpc}\footnotesize
         Dielectric function for $\bE\perp \bchat$.  Solid curve: present model.
Data:
TP65\,=\,\citet{Taft+Philipp_1965},
GHBT69=\citet{Greenaway+Harbeke+Bassani+Tosatti_1969},
TB70=\citet{Tosatti+Bassani_1970},
KSS74=\citet{Klucker+Skibowski+Steinmann_1974},
V75=\citet{Venghaus_1975},
HGK75=\citet{Hagemann+Gudat+Kunz_1974,Hagemann+Gudat+Kunz_1975},
P77=\citet{Philipp_1977},
JHL07=\citet{Jellison+Hunn+Lee_2007},
PP14=\citet{Papoular+Papoular_2014}.
The model $\epsilon_\perp$ plotted here is available
at \website\ and \websiteb\ .}
\end{center}
\end{figure}

There have been numerous studies of $\epsilon_\perp$ from
the far-infrared to soft X-rays (see Table \ref{tab:data}).\footnote{
   Table \ref{tab:data} does not include the studies by
   \citet{Carter+Huebner+Hamm+Birkhoff_1965} or
   \citet{Stagg+Charalampopoulos_1993} because they neglected
   anisotropy.}
Figure \ref{fig:eps Eperpc} shows $\epsilon_\perp$ from various
experimental studies.
It is apparent that there are considerable differences among the
experimental studies.

We adopt the free-electron component parameters obtained by
\citet[][hereafter PP14]{Papoular+Papoular_2014}, who analyzed
reflectivity measurements 
by \citet{Kuzmenko+vanHeumen+Carbone+vanderMarel_2008}
extending to $200\micron$, 
for temperatures $T$ ranging from $10\K$ to $300\K$.
PP14 represented the infrared-optical dielectric function by the sum
of three free-electron-like components,
giving d.c.\ conductivities ranging from
$\sigma_{\rm dc}=1.07\xtimes10^4\mho\cm^{-1}$ at $T=10\K$ to
$1.76\xtimes10^4\mho\cm^{-1}$ at $T=300\K$.
Following PP14, 
we also use three free-electron-like components.  
For components 1 and
2 we adopt the parameters recommended by PP14.
However, free-electron component 3 of PP14 contributed strong
absorption at optical and UV frequencies that do not closely match
the laboratory data.
We choose to fit the lab data by adding additional ``resonant'' components,
and therefore modify the parameters for free-electron component 3:
we reduced $\tau_3^{-1}$ by a factor $0.3$,
and reduced $\omega_{p3}$ by a factor $\sqrt{0.3}$.
This leaves $\sigma_{\rm dc} \propto \omega_p^2\tau$ unaffected,
but reduces the contribution of 
component 3 at optical and UV frequencies, where we use additional
``resonant'' contributions to improve the fit to laboratory data.

Graphite has a narrow optically-active in-plane lattice resonance
at $6.30\micron$
\citep{Nemanich+Lucovsky+Solin_1977,
Jeon+Mahan_2005, 
Manzardo+Cappelluti+vanHeumen+Kuzmenko_2012}.
We adopt the resonance parameters from
\citet{Nemanich+Lucovsky+Solin_1977}.
To generate a dielectric function $\epsilon_\perp(\omega)$
compatible with the
experimental results shown in Figure \ref{fig:eps Eperpc},
we add 10 additional ``resonant'' components to represent
electronic transitions.
The adopted
parameters ($\omega_{0r}$, $S_r$, $\gamma_{r}$)
are listed in Table \ref{tab:parameters for Eperpc}.
Note that the sum over oscillator strengths 
$Z_{\rm eff}(\omega\rightarrow\infty) = \sum_jf_{pj}+\sum_j f_{rj}=3.96$,
consistent with the expected sum rule.
The resulting dielectric function is plotted in 
Figure \ref{fig:eps Eperpc}.

\begin{table}[ht]
\caption{\label{tab:parameters for Eperpc}
Component parameters for $\bE\perp \bchat$}
\begin{center}
{\footnotesize
Free-electron-like component parameters for $\bE\perp \bchat$\\
\begin{tabular}{c c c c c c c c c}
\hline
$T$ & $j$ & $A_j$ & $\hbar\omega_{pj}$ & $\omega_{pj}\tau_{{\rm bulk},j}$ & $\tau_{{\rm bulk},j}$ & $f_{pj}$ &
    $\sigma_{\rm dc}$ & $\sigma_{\rm dc}$\\
(K) &     &       & (eV) &  & (s) & & ($10^{15}\s^{-1}$)&
($10^3\mho\cm^{-1}$) \\
\hline
10 & 1$^a$ &  1 & 0.6179 & 128.7 & $1.37\xtimes10^{-13}$ & 0.00241    & $\ 9.62$ & $10.7$\\
   & 2$^a$ &$-1$& 0.2499 & 52.06 & $1.37\xtimes10^{-13}$ & $-0.00039$ & $-1.57$  &\\
   & 3$^b$ &  1 & 6.247  & 2.082 & $2.19\xtimes10^{-16}$ & 0.2466     & $\ 1.57$ &\\
   & total &    &        &       &                       & 0.2486     & $\ 9.62$ & $10.7$\\
\hline
20 & 1$^a$ &  1 & 0.6174 & 128.6 & $1.37\xtimes10^{-13}$ & 0.00241    & $\ 9.60$ & $10.7$\\
   & 2$^a$ &$-1$& 0.2498 & 52.04 & $1.37\xtimes10^{-13}$ & $-0.00039$ & $-1.57$  &\\
   & 3$^b$ &  1 & 6.245  & 2.082 & $2.19\xtimes10^{-16}$ & $0.2465$   & $\ 1.57$ &\\
   & total &    &        &       &                       & 0.2485     & $\ 9.60$ & $10.7$\\
\hline
30 & 1$^a$ &  1 & 0.6137 & 136.4 & $1.46\xtimes10^{-13}$ & 0.00238    & $10.12$  & $11.3$\\
   & 2$^a$ &$-1$& 0.2413 & 53.63 & $1.46\xtimes10^{-13}$ & $-0.00037$ & $-1.56$  &\\
   & 3$^b$ &  1 & 6.231  & 2.077 & $2.19\xtimes10^{-16}$ & 0.2454     & $\ 1.56$ &\\
   & total &    &        &       &                       & 0.2474     & $10.12$  & $11.3$\\
\hline
300& 1$^a$ &  1 & 0.9441 & 138.8 & $9.68\xtimes10^{-14}$ & 0.00563    & $15.85$  & $17.6$\\
   & 2$^a$ &$-1$& 0.9360 & 14.93 & $1.05\xtimes10^{-14}$ & $-0.00554$ & $-1.69$  &\\
   & 3$^b$ &  1 & 6.474  & 2.158 & $2.19\xtimes10^{-16}$ & 0.2649     & $\ 1.69$ &\\
   & total &    &        &       &                       & 0.2650     & $15.85$  & $17.6$\\
\hline
\multicolumn{9}{l}{$^a$ Parameters from \citet{Papoular+Papoular_2014}}\\
\multicolumn{9}{l}{$^b$ $\sigma_{\rm dc}$ from \citet{Papoular+Papoular_2014},
except $\omega_{p3}$ reduced by factor $\sqrt{0.3}$, and
$\tau_3^{-1}$ reduced by }\\ 
\multicolumn{9}{l}{\hspace*{1em}factor $0.3$.}\\
\end{tabular}\\
\bigskip
Resonance-like component parameters for $\bE\perp c$\\
\begin{tabular}{c c c c c l}
\hline
$j$ & $\hbar\omega_{rj}$ & $S_{rj}$ & $\gamma_{rj}$ &$f_{rj}$& note\\
    &  (eV)              &       &               && \\
\hline
1  & 0.1968 & 0.031   & 0.003 & 0.0000 & \citet{Nemanich+Lucovsky+Solin_1977}\\
2  & 2.90   & 2.50    & 0.900 & 0.1328 &\\
3  & 4.40   & 2.20    & 0.230 & 0.2691 &\\
4  & 12.6   & 0.35    & 0.130 & 0.3511 &\\
5  & 14.0   & 0.70    & 0.130 & 0.8669 &\\
6  & 18.0   & 0.17    & 0.350 & 0.3480 &\\
7  & 21.0   & 0.10    & 0.350 & 0.2786 &\\
8  & 31.0   & 0.12    & 0.60  & 0.7286 &\\
9  & 50.0   & 0.036   & 0.70  & 0.5687 &\\
10 & 100.   & 0.003   & 0.70  & 0.1896 &\\
11 & 200.   & 0.0001  & 0.70  & 0.0253 &\\
\hline
   &        &         &       & 3.958  & total from L shell resonances\\
\hline
   &        &         &       & 2.00   & total from K shell\\
\hline
   &        &         &       & 5.958  & total\\
\hline
\end{tabular}
}
\end{center}
\end{table}

\subsection{\label{sec:E para c: IR to EUV}
         $\bE\parallel \bchat$}

\begin{table}[ht]
\caption{\label{tab:parameters for Eparac}
Component parameters for $\bE\parallel c$}
\begin{center}
{\footnotesize
Free-electron-like parameters for $\bE\parallel \bchat$\\
\begin{tabular}{c c c c c c c c}
\hline
$j$ & $A_j$ & $\hbar\omega_{pj}$ & $\omega_{pj}\tau_{{\rm bulk},j}$ & $\tau_{{\rm bulk},j}$ & $f_{pj}$ & $\sigma_{\rm dc}$ &$\sigma_{\rm dc}$\\
    &       & (eV)            &  & (s) & & $(10^{12}\s^{-1}$) & 
$\mho{\cm}^{-1}$\\
\hline
1 & 1 & 0.25 & 0.38 & $1.00\xtimes10^{-15}$ & 0.000394 & $11.5$ & 12.8\ \\
\hline
\end{tabular}\\
\bigskip
Resonance parameters for $\bE\parallel \bchat$\\
\begin{tabular}{c c c c c l}
\hline
$j$ & $\hbar\omega_{rj}$ & $S_{rj}$ & $\gamma_{rj}$ &$f_{rj}$& note\\
    &  (eV)              &       &               && \\
\hline
1 & 0.1075 & 0.004 & 0.001 & 0.0000 & \citet{Nemanich+Lucovsky+Solin_1977}\\
2 & 4.40   & 1.4   & 0.480 & 0.1713  &\\
3 & 7.20  & 0.045  & 0.190 & 0.0147 &\\
4 & 11.1  & 0.95   & 0.120 & 0.7396  &\\
5 & 12.1  & 0.21   & 0.110 & 0.1943  &\\
6 & 13.5  & 0.23   & 0.200 & 0.2649  &\\
7 & 16.3  & 0.26   & 0.210 & 0.4365  &\\
8 & 22.5  & 0.20   & 0.350 & 0.6397  &\\
9& 31.0  & 0.12   & 0.600 & 0.7286  &\\
10& 50.0  & 0.036  & 0.700 & 0.5687  &\\
11& 100.  & 0.003  & 0.700 & 0.1896  &\\
12& 200.  & 0.0001 & 0.700 & 0.02527 &\\
\hline
  &       &         &      & 3.976  & total from L shell resonances\\
\hline
  &       &         &      & 2.00   & total from K shell\\
\hline
  &       &         &      & 5.976  & total\\
\hline
\end{tabular}
}
\end{center}
\end{table}
\begin{figure}[h]
\begin{center}
\includegraphics[angle=0,width=8.0cm,
                 clip=true,trim=0.5cm 0.5cm 0.5cm 0.5cm]
{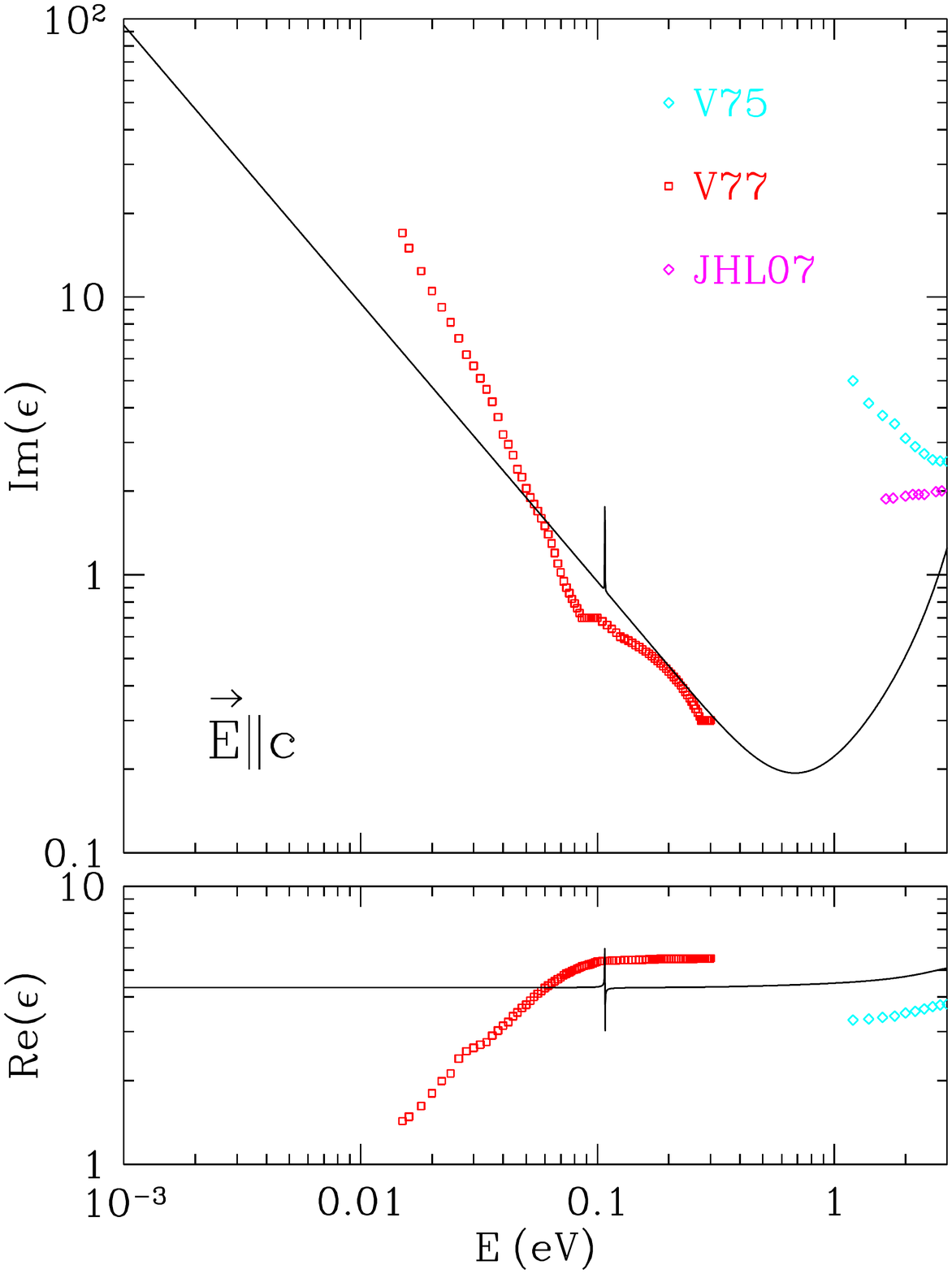}
\includegraphics[angle=0,width=8.0cm,
                 clip=true,trim=0.5cm 0.5cm 0.5cm 0.5cm]
{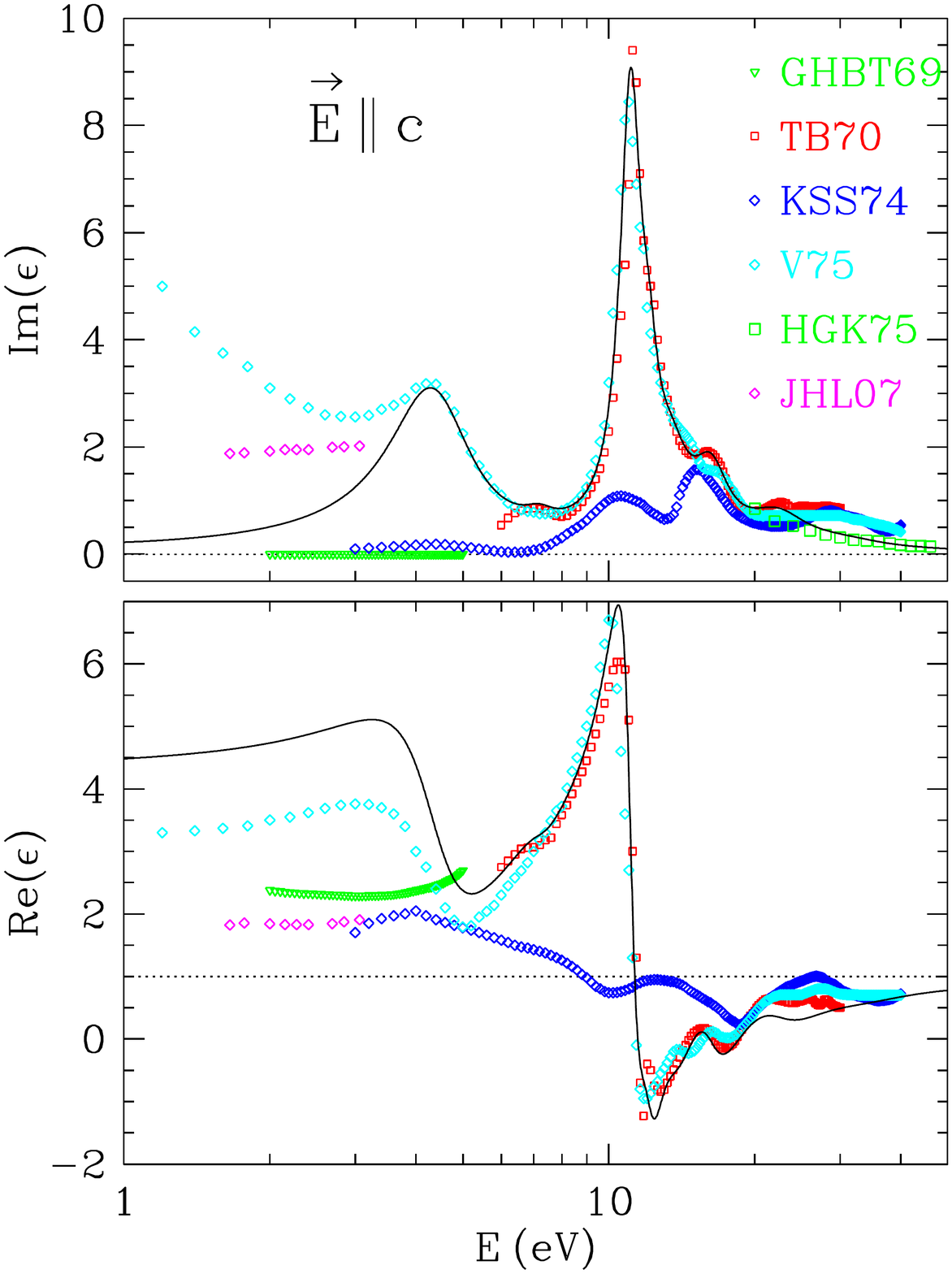}
\caption{\footnotesize \label{fig:epsilon_para}
Dielectric function for $\bE\parallel \bchat$.  Solid curve: present model.
Data:
GHBT69\,=\,\citet{Greenaway+Harbeke+Bassani+Tosatti_1969},
TB70\,=\,\citet{Tosatti+Bassani_1970},
KSS74\,=\,\citet{Klucker+Skibowski+Steinmann_1974},
HGK75\,=\,\citet{Hagemann+Gudat+Kunz_1974,Hagemann+Gudat+Kunz_1975},
V75\,=\,\citet{Venghaus_1975},
V77\,=\,\citet{Venghaus_1977},
JHL07\,=\,\citet{Jellison+Hunn+Lee_2007}.
The model $\epsilon_\parallel$ plotted here is available
at \website\ and \websiteb\ .}
\end{center}
\end{figure}

The dielectric function for $\bE\parallel \bchat$ is much more uncertain
than that for $\bE\perp \bchat$.
The uncertainty is attributable in part to experimental
difficulties, but is likely also due to real sample-to-sample
variations, particularly as regards the weak conduction resulting from
electrons or holes transiting from one graphene sheet to
another.

We adopt free-electron component parameters corresponding to
$\sigma_{\rm dc}=12\mho\cm^{-1}$, intermediate between 
$\sim$$1\mho\cm^{-1}$ \citep{Klein_1962}
and $\sim$$200\mho\cm^{-1}$ \citep{Primak_1956}
measured for high-quality crystals at $T=300\K$.
With a mean-free-time $\tau_{\rm bulk}=1\times10^{-15}\s$,
corresponding to a mean-free-path $v_{\rm F}\tau_{\rm bulk}\approx 4\Angstrom$
(approximately the interplane spacing),
we obtain a free-electron contribution 
as shown in Figure \ref{fig:epsilon_para}.
The adopted free-electron parameters are approximately consistent
with the data of \citet{Venghaus_1977}, which appears
to be the only study of $\epsilon_\parallel$
in the far-infrared and mid-infrared (see Figure \ref{fig:epsilon_para}).

Graphite has an optically-active out-of-plane lattice resonance at 
$11.52\micron$
\citep{
Nemanich+Lucovsky+Solin_1977,
Jeon+Mahan_2005}.
We adopt the resonance parameters from \citet{Nemanich+Lucovsky+Solin_1977}.

Surprisingly, 
there do not appear to be published experimental determinations of
$\epsilon_\parallel$ between 0.3 and 1\,eV.
At optical frequencies, some studies 
\citep{Greenaway+Harbeke+Bassani+Tosatti_1969,Klucker+Skibowski+Steinmann_1974}
find negligible absorption 
[${\rm Im}(\epsilon_\parallel)\approx 0$] for
$2 < h\nu < 4\eV$, 
while other investigators \citep{Venghaus_1975,Jellison+Hunn+Lee_2007}
report strong absorption
[${\rm Im}(\epsilon_\parallel)\gtsim 1$] 
at these energies.
Our adopted $\epsilon_\parallel$, using 12 resonance-like components
(parameters listed in Table \ref{tab:parameters for Eparac}),
has
moderately strong absorption in the optical and near-IR,
and is reasonably consistent with lab data at higher energies
(see Figure \ref{fig:epsilon_para}).
Note that 
$Z_{\rm eff}(\omega\rightarrow\infty) = \sum_jf_{pj}+\sum_j f_{rj}=3.98$,
consistent with the expected sum rule.

\section{\label{sec:Kshell}
         K Shell Absorption}

Absorption by the $1s^2$ shell in graphite is also
polarization-dependent.
Using polarized synchrotron radiation,
\citet{Rosenberg+Love+Rehn_1986} measured absorption in a HOPG sample
at 7 inclinations of the $c$-axis relative 
to the polarization.
The quantity measured was the electron yield $Y$, including
photoelectrons, Auger electrons, and secondary electrons emitted following
absorption of an X-ray photon; after
subtraction of a ``baseline'' contributed by absorption by the
L shell ($2s^22p^2$) electrons, $Y$ is
assumed to be proportional to the absorption coefficient contributed
by the K shell.
Similar measurements of polarization-dependent absorption near the K edge
in graphene have also been reported
\citep{Pacile+Papagno+Rodriguez+etal_2008,
       Papagno+FraileRodrigues+Girit+etal_2009}.

\begin{figure}[ht]
\begin{center}
\includegraphics[angle=0,width=8.2cm,
                 clip=true,trim=0.5cm 1.0cm 1.3cm 0.5cm]
{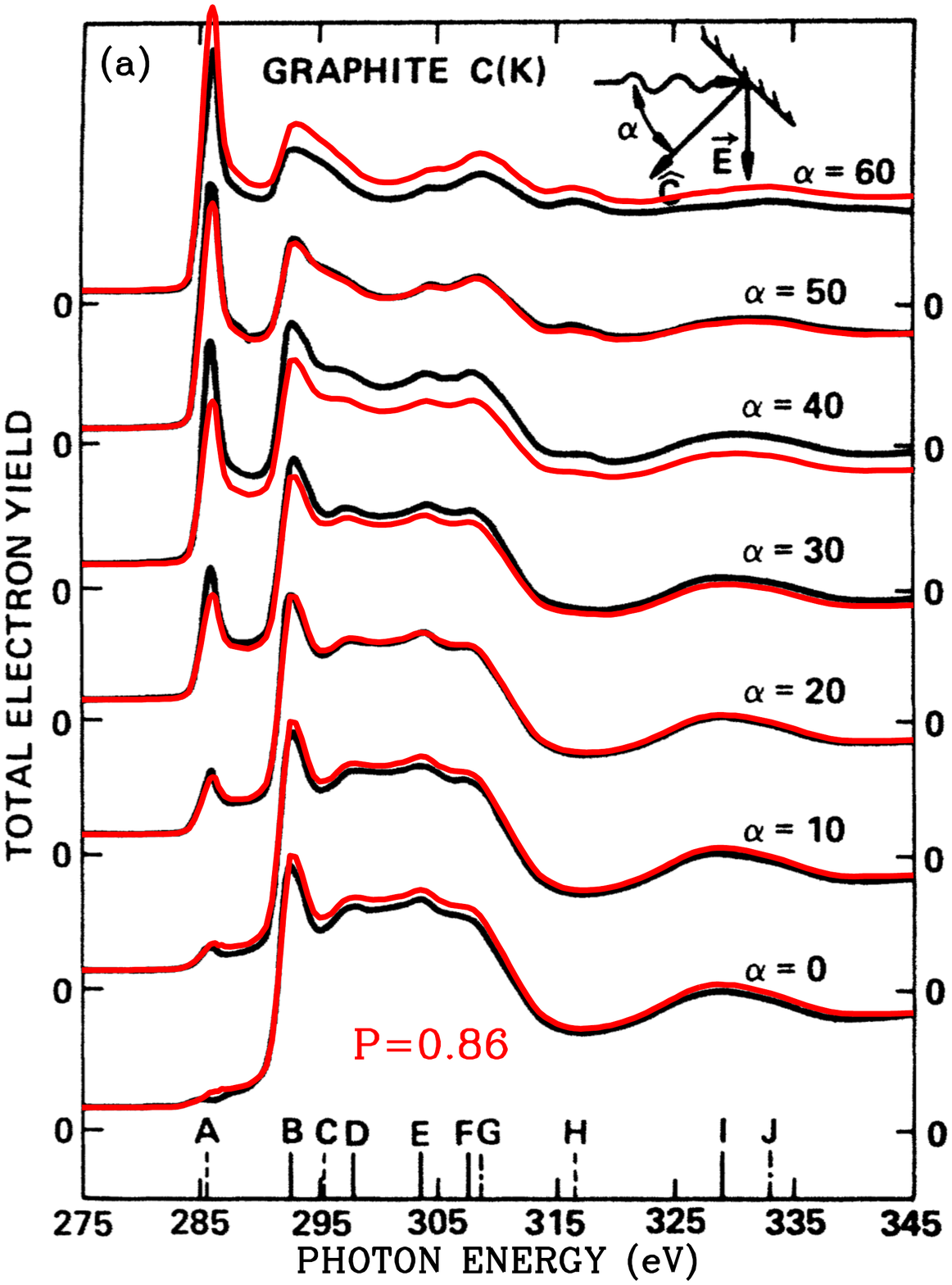}
\includegraphics[angle=0,width=8.1cm,
                 clip=true,trim=0.5cm 0.5cm 0.5cm 0.5cm]
{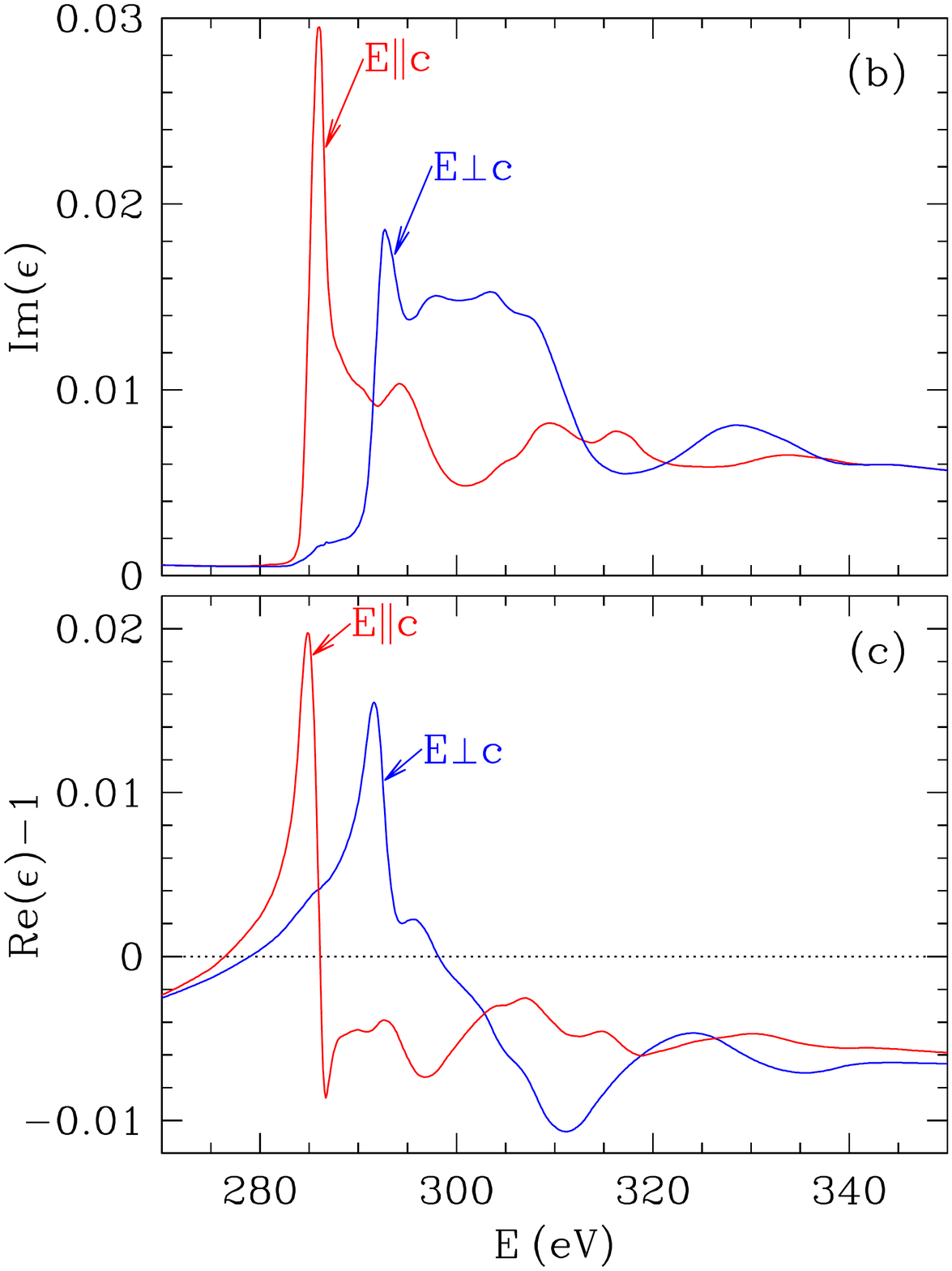}
\caption{\label{fig:eps2K}\footnotesize
The near-edge X-ray absorption fine structure in graphite has a
pronounced dependence on incident polarization.
(a) Electron yield $Y(E,\alpha)$ from \citet{Rosenberg+Love+Rehn_1986}
with our model (red).
(b, c): Our estimate for
$\epsilon_\perp$ and
$\epsilon_\parallel$.
$\epsilon_\perp$ and $\epsilon_\parallel$ are
available at \website\ and \websiteb\ .}
\end{center}
\end{figure}
\begin{figure}[ht]
\begin{center}
\includegraphics[angle=0,width=10.0cm,
                 clip=true,trim=0.5cm 0.5cm 0.5cm 0.5cm]
{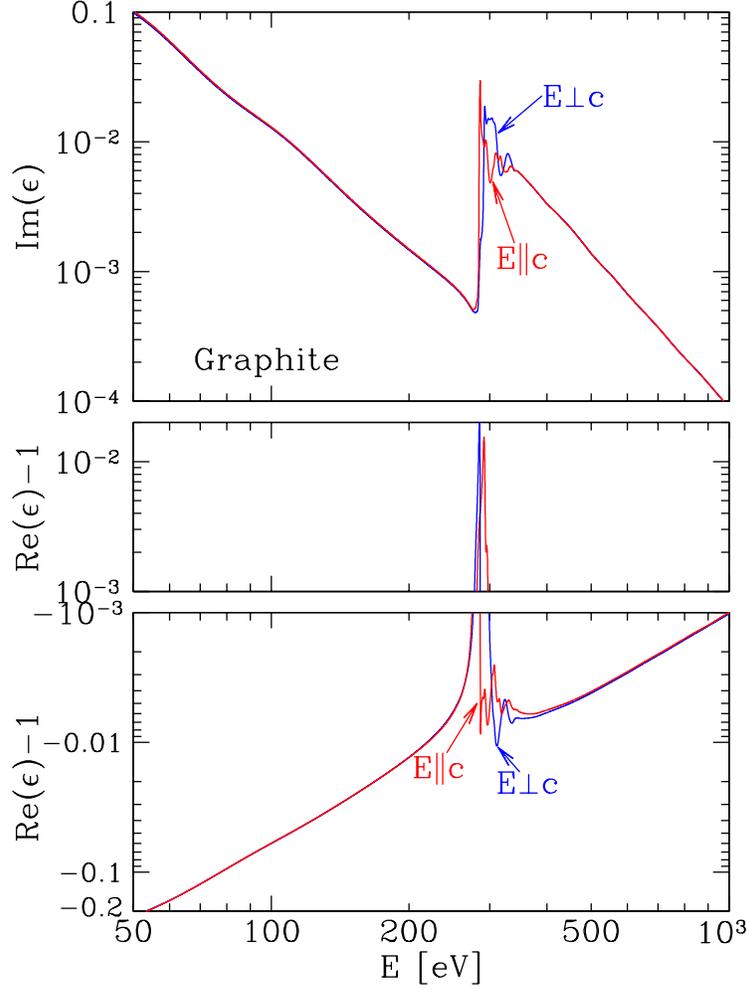}
\caption{\label{fig:epsuvxray}\footnotesize
Dielectric tensor for graphite from $50\eV$ to $700\eV$ (see text).
From $50$ to $280\eV$ the absorption is due to photoionization from
the $2s^22p^2$ shell.  Above $280\eV$ the absorption is dominated
by the electrons in the K shell.
$\epsilon_\perp$ and $\epsilon_\parallel$ are
available at \website\ and \websiteb\ .}
\end{center}
\end{figure}
Consider X-rays incident on the graphite basal plane,
with $\alpha$ the angle between the incident direction
${\bf k}_{\rm in}$ and the surface normal $\bchat$, and 
assume the incident radiation to have fractional polarization $P$ in 
the ${\bf k}_{\rm in}-\bchat$ plane.
If $Y_\perp(E)$ and $Y_\parallel(E)$ are the yields
for incident $\bE\perp\bchat$ and $\bE\parallel\bchat$,
then the yield is
the appropriate weighted average of $Y_\perp$ and $Y_\parallel$:
\beq
Y(E,\alpha)=\left[\frac{(1+P)}{2}\cos^2\alpha+\frac{(1-P)}{2}\right]Y_\perp(E)
+
\left[\frac{(1+P)}{2}\sin^2\alpha\right] Y_\parallel(E)
\eeq
where the quantities in square brackets are the fractions of the
incident power with $\bE\perp\bchat$ and $\bE\parallel\bchat$.
$Y(E,\alpha)$ from
\citet{Rosenberg+Love+Rehn_1986} were used to infer
$Y_\perp(E)$ and $Y_\parallel(E)$ (arbitrary units) for $280\leq E\leq 345\eV$
(see Figure \ref{fig:eps2K}a), assuming the incident
radiation to have polarization fraction $P=0.86$ \citep{Stohr+Jaeger_1982}.
To obtain the contribution $\delta\epsilon_2$ of
the K shell electrons to $\epsilon_2$, we assume
that $\omega\delta\epsilon_2 \propto \delta Y(E)$ where $\delta Y$
is the contribution of the K shell.
For $E<345\eV$ we inferred $\delta Y_\perp, \delta Y_\parallel$ from the
data of \citet{Rosenberg+Love+Rehn_1986}, and
set 
\beq
{\rm Im}\left(\delta\epsilon_{\rm K}\right) = A \frac{\delta Y(E)}{E}
~~~.
\eeq
For $E>345\eV$ we assumed ${\rm Im}(\delta\epsilon_{\rm K})\propto E^{-4}$
(i.e., absorption coefficient $\propto E^{-3}$).
To determine the constant $A$ we
require that $\delta\epsilon_2$ obey the sum
rule \citep{Altarelli+Dexter+Nussenzveig+Smith_1972}
\beq
\int_0^\infty \omega \frac{{\rm Im}(\delta\epsilon_{{\rm K}\parallel} + 
2\delta\epsilon_{{\rm K}\perp})}{3} 
d\omega =
\frac{2\pi^2 e^2}{m_e} Z_K n_{\rm C}
~~~,
\eeq
where $Z_K=2$ is the number of K shell electrons per C,
and $n_{\rm C}$ is the number density of
C atoms. 
${\rm Re}(\delta\epsilon_{\rm K})$ is obtained from 
${\rm Im}(\delta\epsilon_{\rm K})$ using the Kramers-Kronig relation
(\ref{eq:KK}).
Combining $\delta\epsilon_K(\omega)$ with the free-electron and
resonance model (see Eq.\ \ref{eq:epsilon}) yields 
a dielectric tensor for graphite extending continuously
from $0$ to $10\keV$.  
The resulting $\epsilon_{\perp}$ and $\epsilon_{\parallel}$ are
shown in Figure \ref{fig:eps2K}b,c.
The near-edge absorption is strongly polarization-dependent: 
For $\bE\parallel\bchat$ the absorption peaks at 286$\eV$,
whereas for $\bE\perp\bchat$ the absorption peaks at 293$\eV$.

Figure \ref{fig:epsuvxray} shows the dielectric function from
$50\eV$ to $1\keV$.
Between $50\eV$ and $280\eV$ the absorption is smooth and featureless,
arising from photoionization of the 4 electrons in the L shell.
At $\sim$$282\eV$ the 
onset of K shell absorption causes ${\rm Im}(\epsilon)$
to increase by a factor $\sim$50.
For energies $E\gtsim350\eV$, the photolectric 
opacity declines smoothly
with increasing energy.

\begin{figure}[ht]
\begin{center}
\includegraphics[angle=270,width=10.0cm,
                 clip=true,trim=0.5cm 0.5cm 0.5cm 0.5cm]
{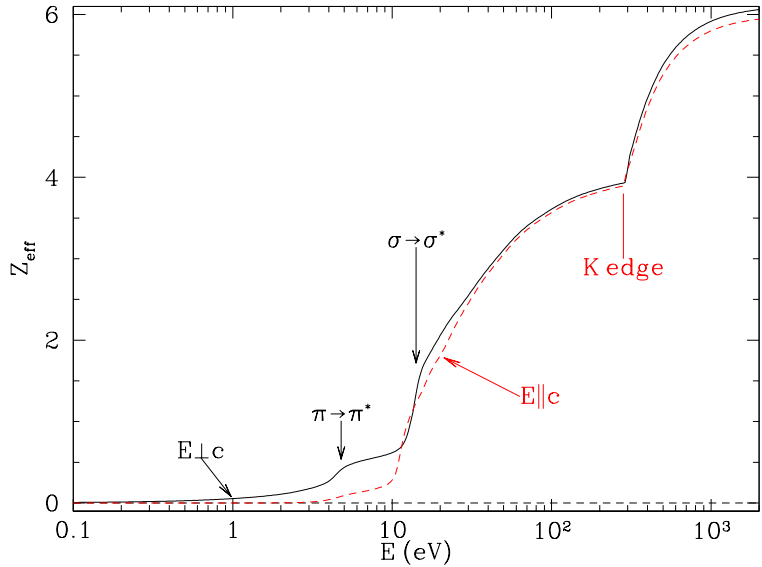}
\caption{\footnotesize \label{fig:neff}
         $Z_{\rm eff}(E)$, the effective number of electrons per C 
         producing absorption at $\hbar\omega\leq E$
         (see Eq.\ \ref{eq:neff})
         for $\epsilon_{\perp}$ and $\epsilon_\parallel$.
        }
\end{center}
\end{figure}

Figure \ref{fig:neff}
shows the effective number $n_{\rm eff}(E)$
of electrons contributing to absorption at $\hbar\omega\leq E$
for $\epsilon_{\perp}$ and $\epsilon_{\parallel}$.

\section{\label{sec:cross sections}
         Absorption and Scattering by Single-Crystal Spheres and Spheroids}

Consider a spheroid with semi-axes $(a,b,b)$,
composed of a uniaxial material such as graphite.
Assume $\bchat\parallel\bahat$, where $\bchat$ is the crystal $c$-axis
and $\bahat$ is the spheroid symmetry axis.
The volume $V=(4\pi/3)ab^2$, and 
$\aeff\equiv (ab^2)^{1/3}$ is the radius of an equal-volume sphere.

Calculation of absorption and scattering by particles composed of
anisotropic materials is a difficult problem, even for spheres.  
Accurate calculation of $C_\abs$
and $C_\sca$ requires
solving Maxwell's equations for an incident plane wave interacting
with the target, using a method
that can explicitly treat anisotropic materials.  While
certain approximations (see \S\ref{sec:onethirdtwothird})
can be used when $\aeff/\lambda\ll 1$,
analytic treatments of the general case are lacking,
and we are forced to rely on numerical methods.

\subsection{\label{sec:DDA}
            Accurate Results: the Discrete Dipole Approximation}

The discrete dipole approximation (DDA)
\citep{Purcell+Pennypacker_1973,
       Draine_1988,Draine+Flatau_1994}
can explicitly treat anisotropic dielectric tensors and 
nonspherical target geometries.
Anisotropic dielectric tensors are treated by assigning 
to each dipole a polarizability tensor with the polarization
response depending
on the direction of the local electric field relative to the local
crystalline axes.
The DDA results are expected to converge to the exact solution to
Maxwell's equations in the limit $N\rightarrow\infty$, where $N$ is
the number of dipoles used to represent the target.
\citet{Draine+Malhotra_1993} used the DDA to show
that the \onethirdtwothird\
approximation was reasonably accurate for $a\ltsim0.04\micron$
graphite spheres
in the vacuum ultraviolet.
However, validity of the \onethirdtwothird\ approximation
for larger particles, or
at optical and near-infrared
wavelengths, does not appear to thus far have been investigated.

We use the DDA code DDSCAT 7.3.1\footnote{
   Available from http://www.ddscat.org}
to calculate scattering and absorption by randomly-oriented
graphite spheres and spheroids of various sizes.
For spheroids, we assume $\bchat\parallel\bahat$.
Let $\Theta$ be the angle between
the incident ${\bf k}$ and $\bahat$; the angular
average
$\langle C \rangle = \int_0^1 C(\Theta)d\cos\Theta$
is evaluated using Simpson's rule and 5 values of 
$\cos\Theta=0,0.25,0.5,0.75,1$.
For each $\Theta$, we average over incident polarizations.
We obtain dimensionless efficiency factors
$Q_\ext(\lambda)\equiv C_\ext(\lambda)/\pi \aeff^2$.

The DDA is exact in the limit $N\rightarrow\infty$.
For sufficiently large $N$, it is expected that the DDSCAT error will
scale as $N^{-1/3}$.  This scaling law is
expected intuitively
($N^{-1/3} \propto$ the fraction of the dipoles that are
located on the surface of the target)
and has been verified computationally
\citep[see, e.g.][]{Collinge+Draine_2004}.
Thus we expect
\beq \label{eq:dependence on N}
Q_N \approx Q_\infty + A N^{-1/3}
~~~,
\eeq
and we can therefore estimate the ``exact'' result from calculations with
$N=N_1$ and $N_2$:
\beqa \label{eq:extrap}
Q_\infty &\approx& Q_{N_2}\times\left(1-\phi_{N_2}\right)
\\ \label{eq:phi_N}
\phi_{N_2} &\equiv& \frac{Q_{N_1}/Q_{N_2}-1}{(N_2/N_1)^{1/3}-1}
~~~,
\eeqa
and the coefficient $A$ in Eq.\ (\ref{eq:dependence on N}) is
\beq \label{eq:A}
A = \frac{(Q_{N_2}-Q_{N_1})N_2^{1/3}}{1-(N_2/N_1)^{1/3}}
= Q_{N_2}\times \phi_{N_2}\times\left(\frac{N_1}{N_2}\right)^{1/3}
~~~.
\eeq
In practice, one chooses as large a value of $N_2$ as is
computationally feasible, and then chooses a value of $N_1\ltsim N_2/2$.
If we simply took $Q_{N_2}$ as our estimate for the
exact result $Q_\infty$, 
the fractional error
would be $\phi_{N_2}$.  Because (\ref{eq:dependence on N}) is
expected to closely describe the dependence on $N$, the extrapolation 
(\ref{eq:extrap}) should yield an estimate for the exact result
$Q_\infty$ with a fractional error much smaller than $\phi_{N_2}$, where
$N_2$ is the largest value of $N$ for which converged DDA results
are available.
As a simple rule-of-thumb, we suggest that $Q_{N_2}\times(1-\phi_{N_2})$
will approximate the ``exact'' result $Q_\infty$ to within a fractional
error $\sim$$0.1\phi_{N_2}$.

\begin{figure}[ht]
\begin{center}
\includegraphics[angle=0,width=8.0cm,
                 clip=true,trim=0.5cm 0.5cm 0.5cm 0.5cm]%
{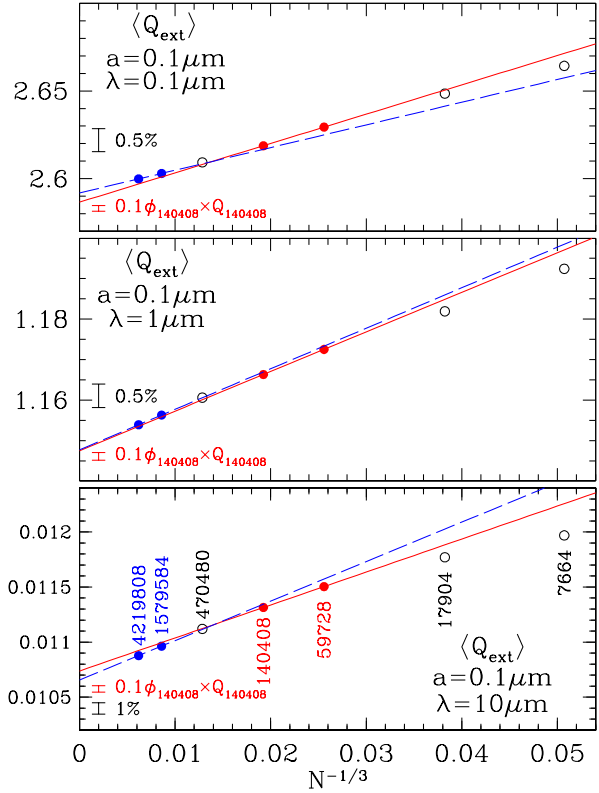}
\caption{\footnotesize \label{fig:dda conv}
         Convergence of DDA results for $\langle Q_\ext\rangle$
         as $N^{-1/3}\rightarrow 0$ for
         $a=0.1\micron$ single-crystal graphite sphere
         and $\lambda=0.1, 1, 10\micron$.
         The red line is 
         Eq.\ (\ref{eq:dependence on N})
         fitted to the results for
         $N_1=59708$ and $N_2=140408$ (red points).
           Red vertical bars show $0.1\phi_{140408}Q_{140408}$,
           the estimated uncertainty in the extrapolation to
           $N=\infty$ from results for $N_1=59708$ and $N_2=140408$.
           The broken blue line is 
           Eq.\ (\ref{eq:dependence on N})
           fitted to the results for $N_1=1.58\times10^6$ and
           $N_2=4.22\times10^6$.
           Black
         vertical bars show shifts of 1\% and 0.5\%.
         Eq.\ (\ref{eq:extrap}) allows reliable extrapolation to
         $N\rightarrow\infty$ provided $N_1,N_2$ are sufficiently
         large, with estimated fractional uncertainty $\sim$$0.1\phi_{N_2}$.
         }
\end{center}
\end{figure}
In Figure \ref{fig:dda conv}
we show results calculated 
for $a=0.1\micron$ graphite spheres at 3 wavelengths: 
$\lambda=10\micron$, $1\micron$, and $0.1\micron$.
For each case the DDA calculations were done with 7 different values of
$N$, ranging from $N\!=\!7664$ to $N\!=\!4.22\xtimes10^6$.
The solid line is
Eq.\ (\ref{eq:dependence on N}) fitted to the results for
$N_1\!=\!59708$ and $N_2\!=\!140408$,
and the dashed line is 
Eq.\ (\ref{eq:dependence on N}) fitted to the results for
$N_1\!=\!1.58\xtimes10^6$ and $N_2\!=\!4.22\xtimes10^6$.
We see that the numerical results conform quite well to the
functional form (\ref{eq:dependence on N}), 
with $\sim$$0.1\phi_{N_2}$ providing a reasonable estimate for the
uncertainty in the extrapolation to the exact result.

\begin{figure}[ht]
\begin{center}
\includegraphics[angle=0,width=12.0cm,
                 clip=true,trim=0.5cm 5.0cm 0.5cm 2.5cm]
{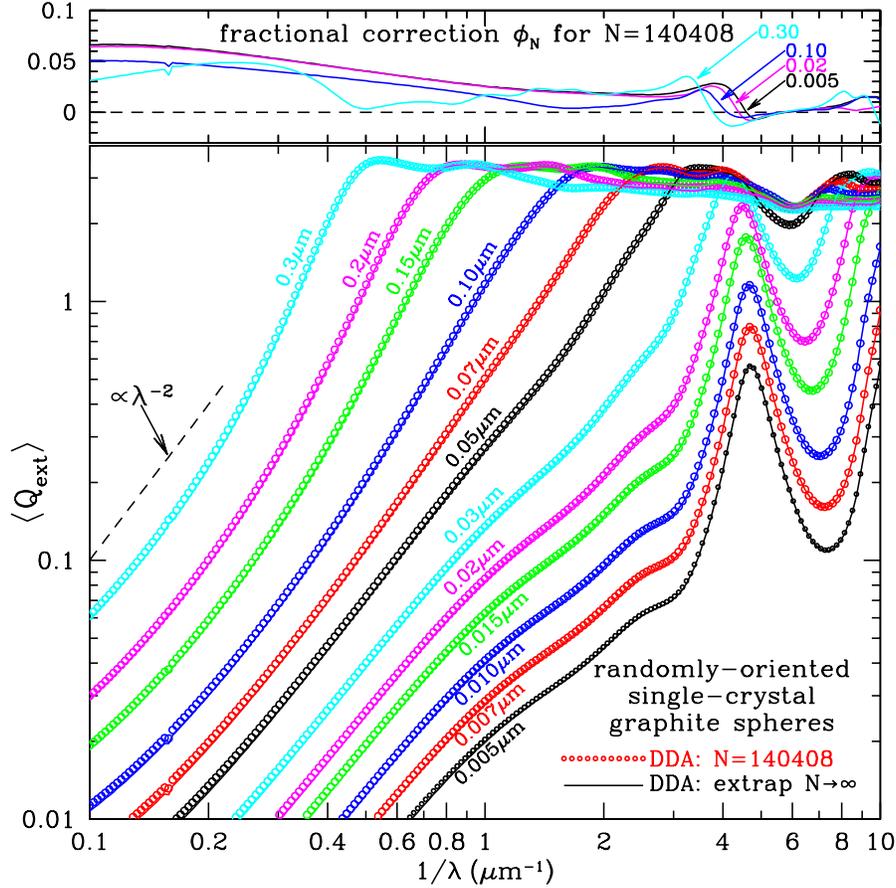}
\caption{\footnotesize \label{fig:dda_sphere_test}
         Lower panel: $\langle Q_\ext(\lambda)\rangle$ for randomly-oriented
         graphite spheres, calculated with the DDA with
         the anisotropic dielectric tensor of graphite, for
         radii from $0.005\micron$ to $0.3\micron$.
         DDA results are shown for
         $N=140408$ dipoles (circles) and the extrapolation
         to $N\rightarrow\infty$ using results for $N_1=59728$
         and $N_2=140408$
         (solid curves).
         Upper panel: the fractional difference $\phi_N$ between the
         DDA results for $N=140408$ and the $N\rightarrow\infty$
         extrapolation, for 4 selected radii (indicated). 
         The fractional uncertainty in the \newtext{DDA}
         extrapolation is
         $\sim$$0.1\phi_N$, i.e., less than 1\%
         for all cases shown.
         DDA cross sections plotted here are available at
         \website\ and \websiteb\ .}
\end{center}
\end{figure}

To survey many combinations of $(a,\lambda)$ 
we consider $0.005\micron\leq\aeff\leq0.3\micron$ and
$10\micron \geq \lambda \geq 0.1\micron$
(scattering parameter $x\equiv 2\pi \aeff/\lambda$
ranging from $0.00314$ to $18.85$). 
Figure \ref{fig:dda_sphere_test} shows fully-anisotropic DDA
calculations of
$\langle Q_\ext\rangle$ for randomly-oriented single-crystal graphite
spheres.  The circles are the DDA results for $N=140408$ dipoles,
and the solid curves are the DDA results extrapolated to $N=\infty$
using Eq.\ (\ref{eq:extrap}).  
The fractional adjustments $\phi_N$ are generally
quite small, so that the curves appear to coincide with the circles.
The upper panel shows $\phi_N$ for 4 different values of $a$,
for $N=140408$.
For all cases shown in Figure \ref{fig:dda_sphere_test},
$|\phi_N|<0.07$ for $N=140408$.

The extrapolation (\ref{eq:extrap}) is expected to provide an
estimate for $Q_\infty$ accurate to a fraction of $\phi_N$; hence
we consider that the extrapolations to $N\rightarrow\infty$ in
Figure \ref{fig:dda_sphere_test} should be accurate to $\sim1\%$ or
better.
The DDA -- which fully allows for anisotropic dielectric tensors --
can therefore be used as the ``gold standard'' to test other, faster,
approximation methods, such as the ``\onethirdtwothird'' approximation.

\subsection{\label{sec:onethirdtwothird}
            The \onethirdtwothird\ Approximation}

If the particle is small compared to the
wavelength, an accurate analytic approximation is available for homogeneous
spheres, spheroids, and ellipsoids, even when composed of
anisotropic materials.
Here we consider spheroids composed of uniaxial material
(such as graphite) with the crystal axis $\bchat$ parallel to the
spheroid symmetry axis $\bahat$.
If $\aeff \ll \lambda\equiv 2\pi c/\omega$, 
the electric dipole approximation \citep{Draine+Lee_1984}
can be used to calculate
absorption and scattering cross sections:
\beqa
C_\abs &=& |\behat\cdot\bahat|^2C_\abs^{\rm (ed)}(\epsilon_\parallel,L_a)
+ \left(1-|\behat\cdot\bahat|^2\right)C_\abs^{\rm (ed)}(\epsilon_\perp,L_b)
\\
C_\sca &=& |\behat\cdot\bahat|^2C_\sca^{\rm (ed)}(\epsilon_\parallel,L_a)
+ \left(1-|\behat\cdot\bahat|^2\right)C_\sca^{\rm (ed)}(\epsilon_\perp,L_b)
~~~,
\eeqa
where $\behat$ is the incident polarization unit vector,
\beqa
C_\abs^{\rm (ed)}(\epsilon,L) &=& \frac{2 \pi V}{\lambda}
\frac{\epsilon_2}{\left|1+(\epsilon-1)L\right|^2}
\\
C_\sca^{\rm (ed)}(\epsilon,L) &=& \frac{8\pi^3 V^2}{3\lambda^4}
\left|\frac{\epsilon-1}{1+(\epsilon-1)L}\right|^2
~~~,
\eeqa
are the ``electric dipole'' cross sections calculated for 
spheroids of volume $V$ with isotropic dielectric function $\epsilon$, and
$L_a$ and $L_b=(1-L_a)/2$ are the usual ``shape factors''\footnote{
  See, e.g., Eq.\ (22.15,22.16) of \citet{Draine_2011a}}
for spheroids.
\begin{figure}[ht]
\begin{center}
\includegraphics[angle=0,width=12.0cm,
                 clip=true,trim=0.5cm 5.0cm 0.5cm 2.5cm]
{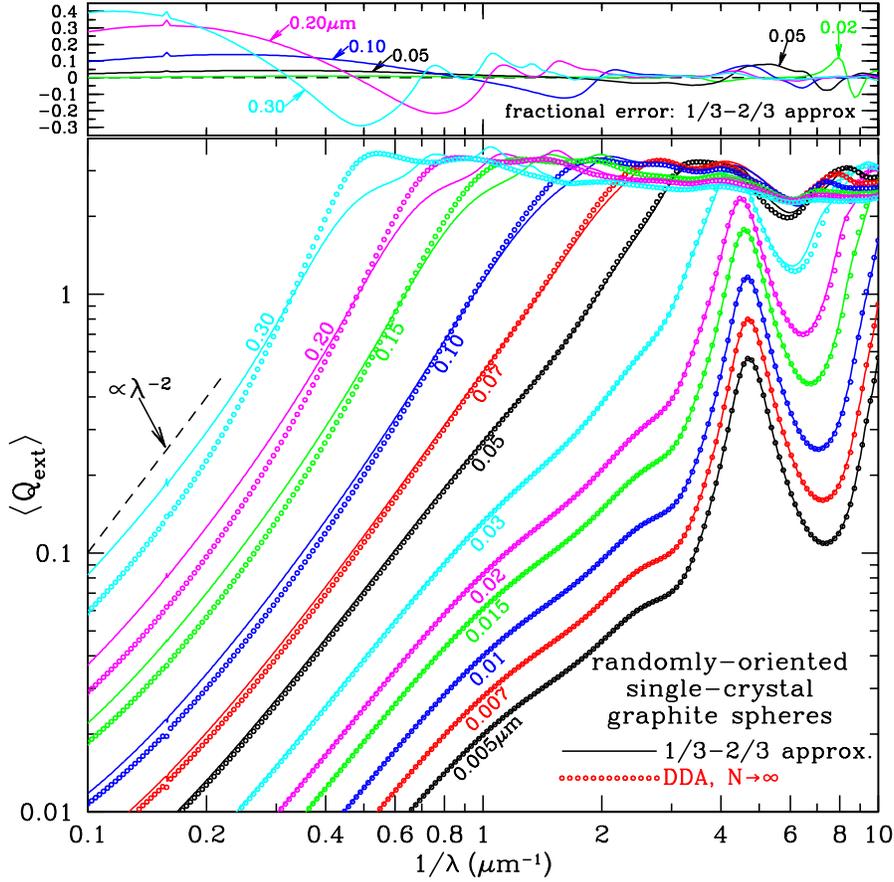}
\caption{\footnotesize \label{fig:dda_sphere}
         Lower panel: $\langle Q_\ext(\lambda)\rangle$ 
         for randomly-oriented single-crystal graphite spheres
         with indicated radii $a$,
         calculated with the \onethirdtwothird\ approximation (solid lines)
         and extrapolated from DDA calculations
         with $N_1=59708$, $N_2=140408$ 
         (open circles; see solid curves in 
         Fig.\ \ref{fig:dda_sphere_test}).
         Upper panel: fractional error resulting from the
         \onethirdtwothird\ approximation.
         For $a=0.3\micron$, the \onethirdtwothird\ approximation overpredicts
         $\langle Q_\ext\rangle$ by 40\% at $\lambda=10\micron$,
         and underpredicts $\langle Q_\ext\rangle$ 
         by 30\% at $\lambda=2\micron$.
         }
\end{center}
\end{figure}

Randomly-oriented spheroids have 
$\langle |\behat\cdot\bahat|^2\rangle=1/3$,
and the average cross section
per particle in the electric dipole limit is simply 
\beqa
\langle C_\abs \rangle
&=& \frac{1}{3} C_\abs^{\rm (ed)} (\epsilon_\parallel,L_a) + 
\frac{2}{3} C_\abs^{\rm (ed)}(\epsilon_\perp,L_b)
\\
\langle C_\sca \rangle
&=& \frac{1}{3} C_\sca^{\rm (ed)} (\epsilon_\parallel,L_a) + 
\frac{2}{3} C_\sca^{\rm (ed)}(\epsilon_\perp,L_b)
~~~.
\eeqa
This is known as the ``\onethirdtwothird\ approximation''.
For crystalline spheres, or for 
crystalline spheroids with $\bahat\parallel\bchat$ or
$\bahat\perp\bchat$, the \onethirdtwothird\ approximation is exact in the 
limit $a/\lambda \rightarrow 0$
because small particles
are then in the ``electric dipole limit'', where the particle's 
response to radiation
can be fully characterized by the induced electric dipole moment
\citep{Draine+Lee_1984}.

The \onethirdtwothird\ approximation,
\beq
\langle C_\abs\rangle \approx \frac{1}{3}C_\abs^{\rm (iso)}(\epsilon_\parallel)
+ \frac{2}{3}C_\abs^{\rm (iso)}(\epsilon_\perp)
\eeq
is
frequently used even when $a/\lambda$ is not small, where now
$C^{\rm (iso)}(\epsilon)$ is a cross section calculated for the same shape
with an {\it isotropic} dielectric function $\epsilon$.
The \onethirdtwothird\ weighting of $C^{\rm (iso)}$
calculated for two different dielectric functions seems plausible
as an approximation, but there have been few tests of its accuracy.

\begin{figure}[ht]
\begin{center}
\includegraphics[angle=0,width=12.0cm,
                 clip=true,trim=0.5cm 5.0cm 0.5cm 2.5cm]
{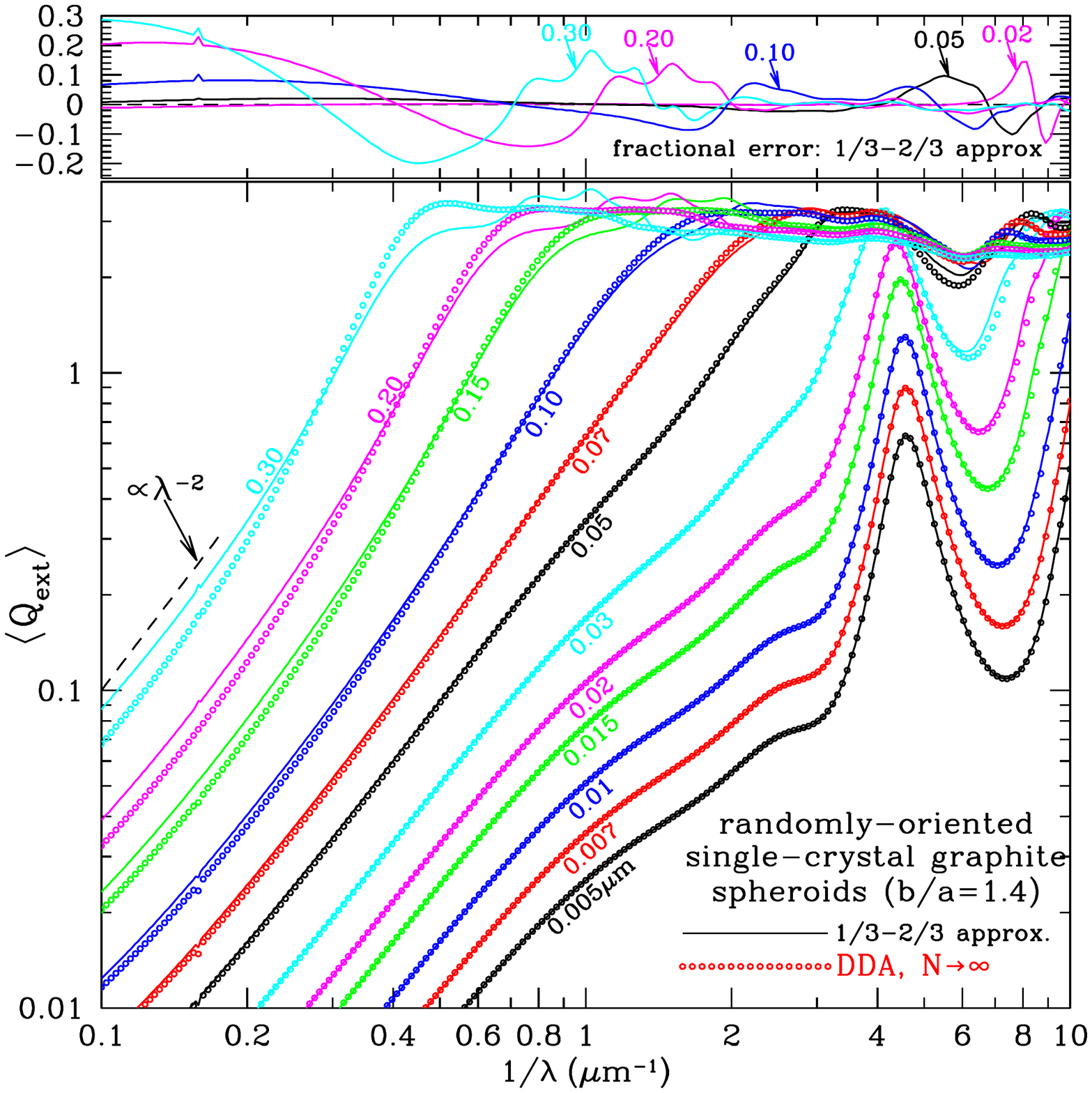}
\caption{\footnotesize \label{fig:dda_spheroid}
         Lower panel:
         $\langle Q_\ext(\lambda)\rangle$ 
         for oblate single-crystal graphite spheroids,
         axial ratio $b/a=1.4$,
         with symmetry axis $\bahat\parallel\bchat$.
         Effective radii $\aeff\equiv(ab^2)^{1/3}$ are indicated.
         Circles: $N\rightarrow\infty$ extrapolation from
         DDA calculations with $N_1=60476$ and $N_2=131040$ dipoles.
         Solid lines: Theory for isotropic spheroids
         \citep{Voshchinnikov+Farafonov_1993}
         using the \onethirdtwothird\ approximation
         (see text).
         Upper panel: fractional error in the \onethirdtwothird\
         approximation for selected radii (indicated).
         DDA cross sections plotted here are available at \website\ 
         and \websiteb\ .}
\end{center}
\end{figure}

Figure \ref{fig:dda_sphere} tests the
\onethirdtwothird\ approximation for spheres.
The points show the ``exact'' results from Fig.\ \ref{fig:dda_sphere_test}
obtained using the DDA
for radii between $0.005\micron$ and $0.3\micron$ and
wavelengths between $10$ and $0.1\micron$.
The solid curves in Figure \ref{fig:dda_sphere} 
show the predictions  of the \onethirdtwothird\ approximation with
Mie theory used to
evaluate $C_\ext^{\rm (iso)}$.
The upper panel in Fig.\ \ref{fig:dda_sphere} shows the
fractional error resulting from use of the \onethirdtwothird\ approximation,
for selected radii.
Over the surveyed domain ($a\leq0.3\micron$, $0.1\leq\lambda\leq10\micron$)
the largest errors occur near $10\micron$ ($E\approx0.1\eV$), 
where $|\epsilon_\perp|$ is
becoming large (see Figure \ref{fig:eps Eperpc}).

As expected, 
the \onethirdtwothird\ approximation is highly accurate 
in the limit $a/\lambda\rightarrow 0$, but
the errors are generally $\ltsim10\%$ in the optical and UV, even
when
$a/\lambda$ is not small.
Relatively large errors occur for $a/\lambda\approx 0.15$ (i.e.,
$2\pi a/\lambda\approx 1$)
where the \onethirdtwothird\ approximation tends to underestimate
$\langle Q_\ext\rangle$.
For example, for $a=0.1\micron$ graphite spheres,
the \onethirdtwothird\ approximation
underestimates $Q_\ext$ by 
12\%
at $\lambda=\newtext{0.6}\micron$ ($\lambda^{-1}=1.5\micron^{-1}$). 
Errors at $a/\lambda\approx 0.15$ are probably of greatest
importance, because for size distributions in the ISM
(e.g., the MRN power law $dn/da \propto a^{-3.5}$, $a\ltsim 0.25\micron$)
and $0.1\micron\ltsim\lambda\ltsim2\micron$,
the grains with
$a\approx 0.15\lambda$ tend to dominate the extinction at
wavelength $\lambda$.
At longer wavelengths, the graphite dielectric tensor becomes large,
especially $\epsilon_\perp$
(see Figs.\ \ref{fig:eps Eperpc} and \ref{fig:epsilon_para}),
\newtext{and the errors for given $a/\lambda$ tend to increase.
For $a/\lambda=0.15$, the \onethirdtwothird\ approximation
underestimates $Q_\ext$ by $\sim$30\% at $\lambda=2\micron$.} 
The \onethirdtwothird\ approximation overestimates $\langle Q_\ext\rangle$
by 40\% for $a=0.30\micron$ at $\lambda=10\micron$, even though
$a/\lambda=0.03$ is small.

Figure \ref{fig:dda_spheroid} shows $\langle Q_\ext\rangle$ for
single-crystal oblate spheroids
with axial ratio $b/a=1.4$.  The graphite $c$-axis is assumed to be
parallel to the symmetry axis of the spheroid.
$\langle C_\ext\rangle$ for
randomly-oriented spheroids is obtained by averaging DDA calculations of
$C_\ext=C_\abs+C_\sca$
for 5 different orientations ($\cos\Theta=0,0.25,0.5,0.75,1$) 
of the grain relative to the
incident direction of propagation $\khat$,
extrapolating to $N\rightarrow\infty$ from results computed
for $N_1=60476$ and $N_2=131040$.

Absorption and scattering cross sections were also calculated using
the spheroid code of \citet{Voshchinnikov+Farafonov_1993} and
the \onethirdtwothird\ approximation, using 3 grain orientations:
\beq
\langle C_\ext\rangle \approx
\frac{1}{3}
C_\ext^{\rm (iso)}(\bkhat\!\parallel\!\bahat,\behat\!\perp\!\bahat,\epsilon_\perp)+
\frac{1}{3}
C_\ext^{\rm (iso)}(\bkhat\!\perp\!\bahat,\behat\!\perp\!\bahat,\epsilon_\perp)+
\frac{1}{3}
C_\ext^{\rm (iso)}(\bkhat\!\perp\!\bahat,\behat\!\parallel\!\bahat,\epsilon_\parallel)
~~~,
\eeq
where $C^{\rm (iso)}(\bkhat,\behat,\epsilon)$ is the cross section
for the same shape target but with an isotropic
dielectric function $\epsilon$.
From Fig.\ \ref{fig:dda_spheroid} we see that for $b/a=1.4$ 
the \citet{Voshchinnikov+Farafonov_1993} spheroid code
plus the \onethirdtwothird\ approximation yields estimates for
$\langle Q_\ext\rangle$ that are accurate to within 15\% for
$\lambda<0.9\micron$. 
The errors come from the \onethirdtwothird\ aproximation, as seen
from the similar errors for spheres in Fig.\ \ref{fig:dda_sphere}.
Once again: 
(1)
the \onethirdtwothird\ approximation tends to
underestimate $C_\ext$ for $\aeff/\lambda\approx 0.15$;
(2) the errors are $\ltsim$10\% in the optical-UV;
(3) for given $a/\lambda$, the fractional errors increase at
long wavelengths, as $\epsilon_\perp$ becomes large.
While accurate DDA computations \newtext{are} preferred, the fact that
the \onethirdtwothird\ approximation is exact for
$a/\lambda\rightarrow0$, and moderately accurate for all $a/\lambda$, 
allows it to be used when DDA computations would be infeasible.

\begin{figure}[ht]
\begin{center}
\includegraphics[angle=0,width=8.0cm,
                 clip=true,trim=0.5cm 0.5cm 0.5cm 0.5cm]
{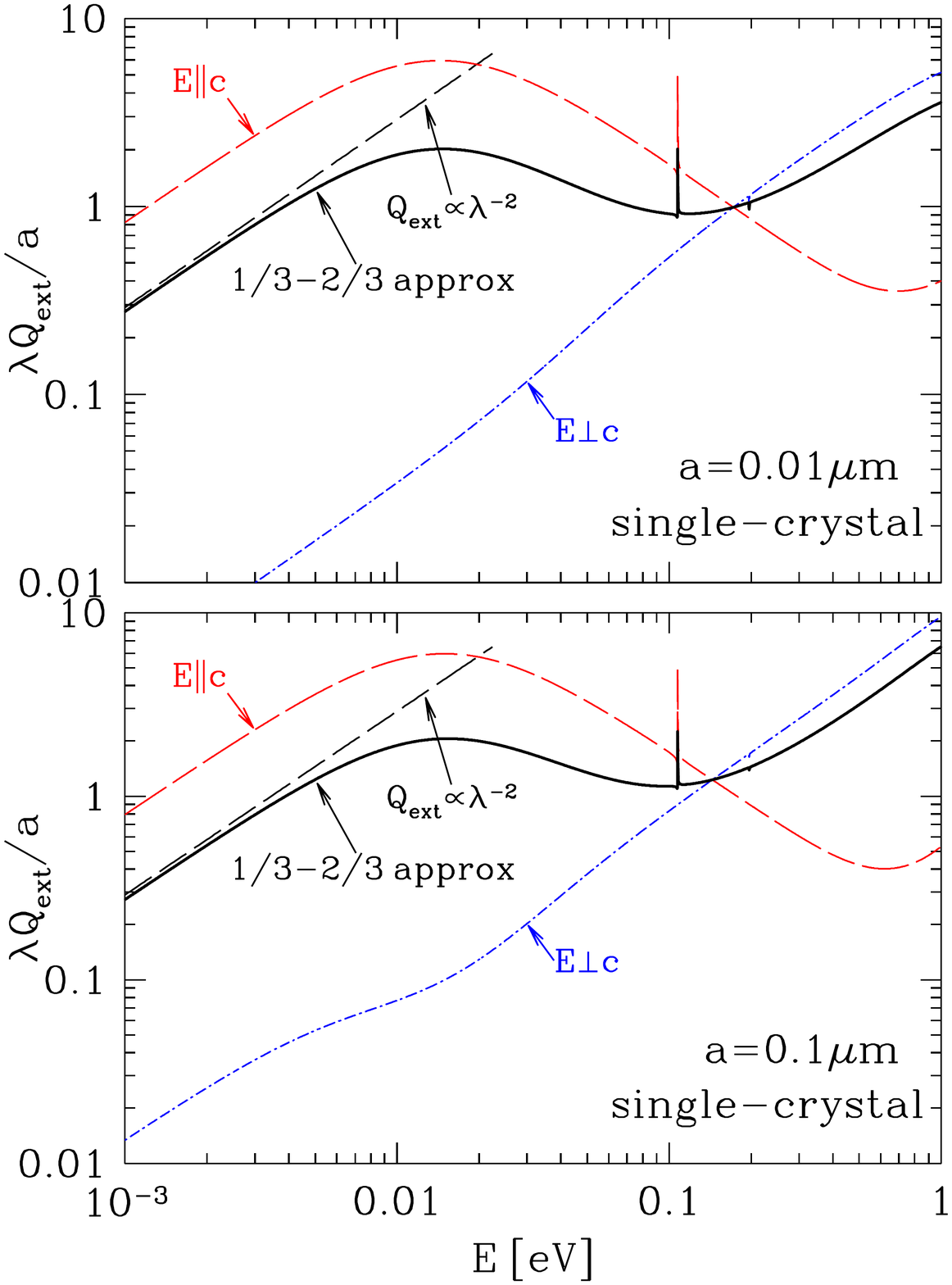}
\includegraphics[angle=0,width=8.0cm,
                 clip=true,trim=0.5cm 0.5cm 0.5cm 0.5cm]
{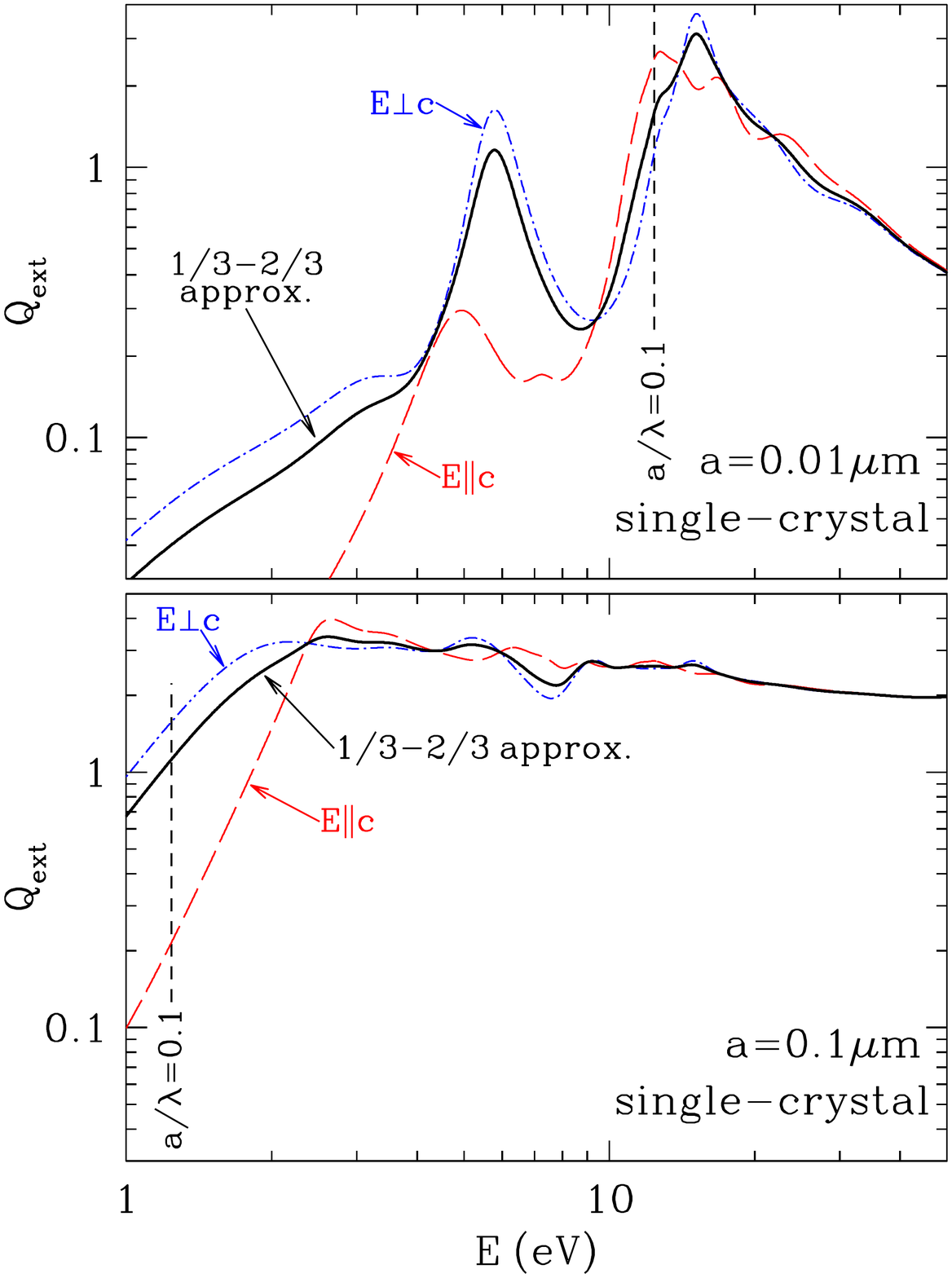}
\caption{\footnotesize \label{fig:Q}
Solid line: $Q_\ext$ for 
randomly-oriented single-crystal graphite spheres,
calculated using Mie theory with the
\onethirdtwothird\ approximation, for 
radii $a=0.01\micron$ and $0.1\micron$.
The \onethirdtwothird\ approximation is expected to be accurate
for $a/\lambda \ltsim 0.1$ (to the left of the vertical dashed lines), 
but for
larger values of $a/\lambda$ its accuracy is uncertain
(but see Fig.\ \ref{fig:dda_sphere}).
Dashed and dash-dot curves are $Q_\ext$ calculated using 
$\epsilon_\perp$ and $\epsilon_\parallel$.
These cross sections, and results for other sizes, are available at
\website\ and \websiteb\ .}
\end{center}
\end{figure}

Having established that the \onethirdtwothird\ approximation 
is moderately accurate for all
$\lambda$, we now use it to calculate cross sections for single-crystal
graphite spheres over a broad range of wavelengths and sizes.
Figure \ref{fig:Q} shows extinction cross sections calculated for
(1) isotropic spheres with $\epsilon=\epsilon_\perp$;
(2) isotropic spheres with $\epsilon=\epsilon_\parallel$;
(3) the \onethirdtwothird\ approximation
(the weighted sum of the above two curves).

We see that for $E\ltsim 0.1\eV$ ($\lambda > 10\micron$), the absorption
is primarily due to the component with $\bE\parallel\bchat$ --  
the conductivity for $\bE\perp\bchat$ is so large that the currents driven
by the applied $\bE$ don't result in as much dissipation as is
associated with the $\bE\parallel\bchat$ component.
For $\lambda \gtsim 200\micron$ the graphite opacity has the ``classical''
$Q_\abs \propto \lambda^{-2}$ behavior.

\section{\label{sec:EMT}
         Turbostratic Graphite and Effective Medium Theory}

``Turbostratic graphite'' refers to
material with the short-range order of graphite, but with randomly-oriented
microcrystallites \citep{Mrozowski_1971,Robertson_1986}.
Highly-aromatized amorphous carbon would be in this category.
\citet[][hereafter PP09]{Papoular+Papoular_2009} argue that turbostratic
graphite (which they refer to as
``polycrystalline graphite'')\footnote{We avoid the term polycrystalline,
because even HOPG is polycrystalline.  Turbostratic graphite consists
of {\it randomly-oriented} microcrystallites.}
in very small ($a\ltsim 0.01\micron$) particles could account for
the observed 2175\AA\ extinction band.
Some of the isotopically-anomalous presolar graphite
grains found in meteorites are composed of turbostratic graphite
\citep{Croat+Stadermann+Bernatowicz_2008,Zinner_2014}.
If interstellar grains contain a carbonaceous component that is highly
aromatic but lacking in long-range order, the optical properties may
be similar to turbostratic graphite.

In principle, the DDA 
can be used to calculate the response from
grains composed of turbostratic graphite, but such calculations
are numerically very challenging because of the need to employ sufficient
dipoles to mimic the arrangement of the microcrystallites, whatever
that may be thought to be.
In the absence of such direct brute-force calculations, one approach is to
employ ``effective medium theory'' (EMT)
to try to
obtain an ``effective'' isotropic dielectric function $\epsilon_{\rm eff}$
for turbostratic graphite, and then calculate scattering and absorption
for homogeneous grains with $\epsilon=\epsilon_{\rm eff}$.

A number of different EMTs have
been proposed \citep[see, e.g.,][]{Bohren+Huffman_1983}.
\begin{figure}[ht]
\begin{center}
\includegraphics[angle=0,width=10.cm,
                 clip=true,trim=0.5cm 0.5cm 0.5cm 0.5cm]
{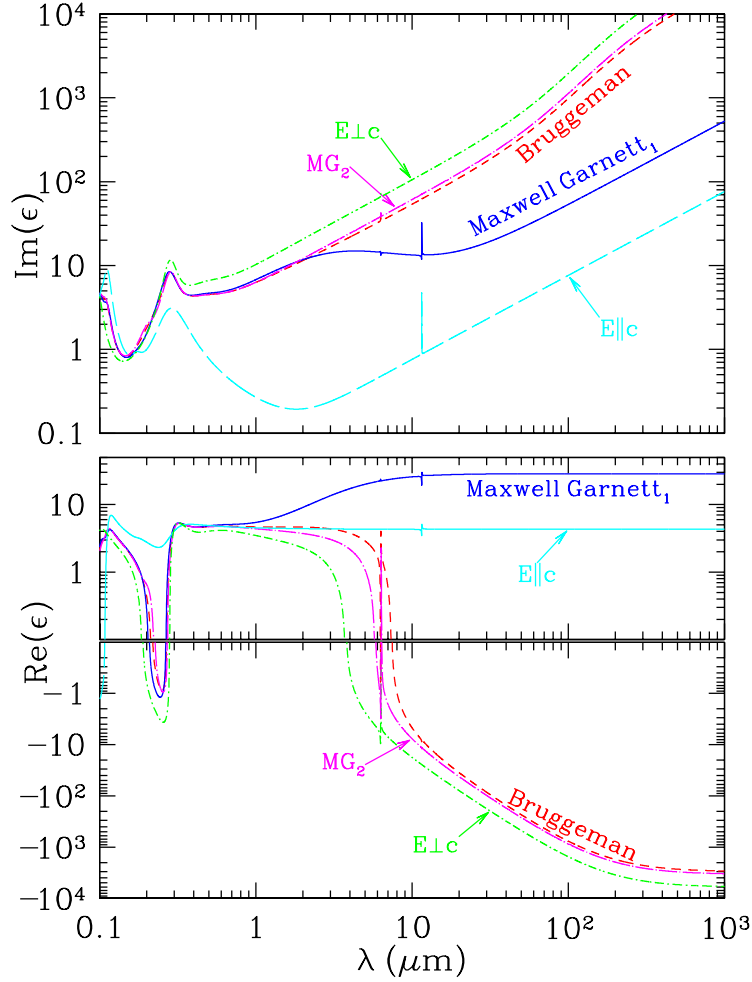}
\caption{\footnotesize\label{fig:eps_EMT}
         Components $\epsilon_\parallel$ and $\epsilon_\perp$ of
         the graphite dielectric tensor, together with effective
         dielectric functions for turbostratic graphite
         obtained from the Maxwell Garnett
         and Bruggeman treatments (see text).
         The MG$_1$ dielectric function is available from
         \website\ and \websiteb\ .}
\end{center}
\end{figure}
\begin{figure}[ht]
\begin{center}
\includegraphics[angle=270,width=12.cm,
                 clip=true,trim=0.5cm 0.5cm 0.5cm 0.5cm]
{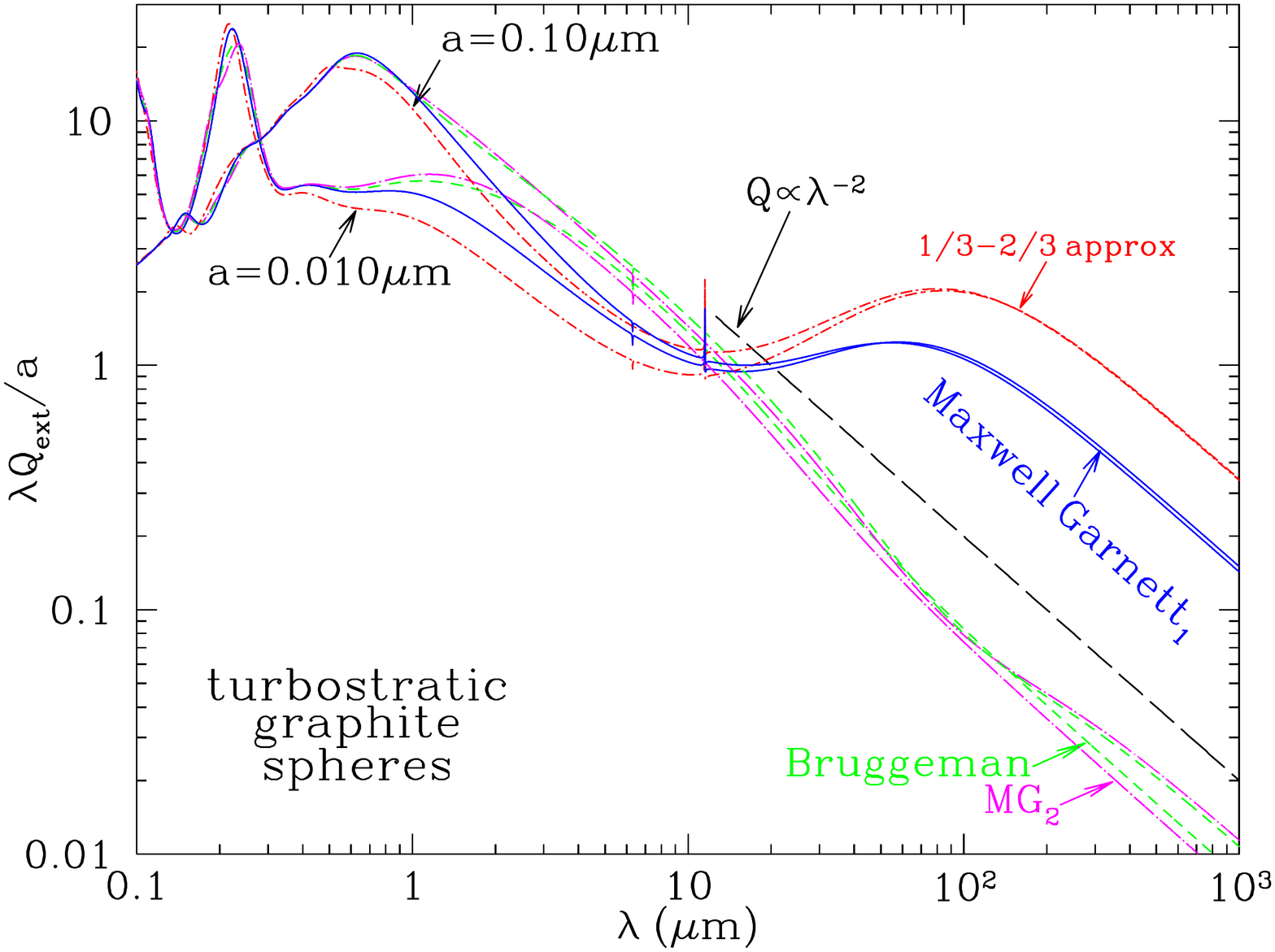}
\caption{\footnotesize\label{fig:EMT}
         Solid curves: 
         $Q_\ext$ for turbostratic graphite grains, estimated
         using Maxwell Garnett EMTs MG$_1$ and MG$_2$,
         for $a=0.01\micron$ and $a=0.10\micron$.
         Also shown are estimates from the
         \onethirdtwothird\ approximation (dot-dashed)
         and the Bruggeman EMT (dashed).
         For $\lambda\ltsim 10\micron$ the \onethirdtwothird\ approximation
         and all EMTs give similar estimates for $Q_\ext$, 
         but for $\lambda>30\micron$ the estimates can differ
         by factors of $\sim$2 or more.
         We recommend the MG$_1$ EMT (see text).
         }
\end{center}
\end{figure}
In the approach of Bruggeman the two components are
treated symmetrically, with
filling factors $\nicefrac{1}{3},\nicefrac{2}{3}$ 
for $\epsilon_\parallel, \epsilon_\perp$.
The Bruggeman estimate $\epsilon_{\rm B}$ 
for the effective dielectric function 
is determined by
\beq
0 = 
\frac{1}{3}
\frac{(\epsilon_\parallel-\epsilon_{\rm B})}
     {(\epsilon_\parallel+2\epsilon_{\rm B})}
+
\frac{2}{3}
\frac{(\epsilon_\perp-\epsilon_{\rm B})}
     {(\epsilon_\perp+2\epsilon_{\rm B})}
~~~,
\eeq
with solution\footnote{In Eq.\ (\ref{eq:EMT}), one must select the
root $r=(\epsilon_\perp^2+8\epsilon_\parallel\epsilon_\perp)^{1/2}$
with ${\rm Im}(r)> 0$.}
\beq \label{eq:EMT}
\epsilon_{\rm B} =
\frac{1}{4}\left[\epsilon_\perp + \left(\epsilon_\perp^2 + 8\epsilon_\parallel\epsilon_\perp\right)^{1/2}\right]
~~~.
\eeq
The Bruggeman EMT was used by PP09 to model turbostratic graphite grains.

The approach of \citet{Maxwell_Garnett_1904} is often used,
where the composite medium is treated as a ``matrix'' 
(with dielectric function
$\epsilon_{\rm matr}$) 
containing inclusions (with dielectric function $\epsilon_{\rm incl}$
and volume filling factor $f_{\rm incl}$.
If the inclusions are taken to be spherical, the 
Maxwell Garnett estimate for the effective dielectric
function is
\beq \label{eq:MG_gen}
\epsilon_{\rm MG} = \epsilon_{\rm matr}
\frac{[3\epsilon_{\rm matr}+(1+2f_{\rm incl})(\epsilon_{\rm incl}-\epsilon_{\rm matr})]}
{[3\epsilon_{\rm matr}+(1-f_{\rm incl})(\epsilon_{\rm incl}-\epsilon_{\rm matr})]}
~~~.
\eeq
Because the two materials are treated asymmetrically, it is necessary
to identify one as the matrix and the other as the inclusion.
As our standard ``Maxwell Garnett EMT'' for turbostratic graphite
we will take $\epsilon_{\rm matr}=\epsilon_\parallel$,
$\epsilon_{\rm incl}=\epsilon_\perp$, and $f_{\rm incl}=2/3$.
Thus,
\beq \label{eq:MG1}
\epsilon_{\rm MG_1} = \epsilon_\parallel
\left[\frac{2\epsilon_\parallel+7\epsilon_\perp}
{8\epsilon_\parallel+\epsilon_\perp}
\right]
~~~.
\eeq
However, we will also examine the consequences of reversing the
choices of matrix and inclusion, and will consider
$\epsilon_{\rm matr}=\epsilon_\perp$, $\epsilon_{\rm incl}=\epsilon_\parallel$,
and $f_{\rm incl}=1/3$:
\beq \label{eq:MG2}
\epsilon_{\rm MG_2} = \epsilon_\perp
\left[\frac{5\epsilon_\parallel+4\epsilon_\perp}
{2\epsilon_\parallel+7\epsilon_\perp}
\right]
~~~.
\eeq
Figure \ref{fig:eps_EMT} shows
$\epsilon_{\rm B}$, $\epsilon_{\rm MG_1}$, and $\epsilon_{\rm MG_2}$ 
derived from
the graphite dielectric tensor components
$\epsilon_\parallel$ and $\epsilon_\perp$.
At wavelengths $\lambda < 1\micron$, $\epsilon_{\rm MG}$,
$\epsilon_{\rm MG_2}$,
and $\epsilon_{\rm B}$ are seen to be nearly identical, 
but at long wavelengths $\epsilon_{\rm MG_1}$ differs
greatly from $\epsilon_{\rm MG_2}$ and $\epsilon_{\rm B}$.
Note that because $\epsilon_{\rm MG_1}$, $\epsilon_{\rm MG_2}$, and
$\epsilon_{\rm B}$ are all analytic functions of
$\epsilon_\parallel$ and $\epsilon_\perp$ 
(see Eqs.\ \ref{eq:EMT}, \ref{eq:MG1} and \ref{eq:MG2}),
it follows \citep[see][]{Bohren_2010} that
$\epsilon_{\rm MG_1}(\omega)$, $\epsilon_{\rm MG_2}(\omega)$, 
and $\epsilon_{\rm B}(\omega)$ 
each satisfies the Kramers-Kronig relations.

\citet{Abeles+Gittleman_1976} found the Maxwell Garnett
EMT to be more successful than the Bruggeman EMT
in characterizing the optical properties
of ``granular metals'', such as sputtered Ag-SiO$_2$ films,
with the insulator SiO$_2$ treated as the ``matrix'' and the Ag
treated as the inclusion, even for Ag filling factors 
$f_{\rm incl}\rightarrow 1$.
This version of the Maxwell Garnett EMT 
was also found to be in better agreement
with measurements on ``granular semiconductors'', such as
Si-SiC, with the stronger absorber (SiC near the SiC stretching
mode at 12.7$\micron$) treated as the inclusion.
This suggests that $\epsilon_{\rm MG_1}$ 
-- which treats the more-strongly-absorbing $\epsilon_\perp$
component as the inclusion -- 
may be the better estimator for turbostratic graphite.

Figure \ref{fig:EMT} shows cross sections for turbostratic
graphite spheres calculated using these \newtext{three} different EMTs.
For comparison, $\langle Q_{\rm ext}\rangle$ is also shown for
randomly-oriented single-crystal spheres, calculated using the
\onethirdtwothird\ approximation.
For $\lambda \ltsim 1\micron$ the \onethirdtwothird\ approximation and
the two EMT variants all 
give comparable estimates for $\langle Q_{\rm ext}\rangle$.
However, for $\lambda\gtsim20\micron$ the estimates diverge.

For turbostratic graphite, we might expect the absorption at long wavelengths
to be less than given by the \onethirdtwothird\ approximation, because
the high conductivity for $\bE\perp\bchat$ allows it to ``screen''
the regions characterized by $\epsilon_\parallel$
from the electric field of the incident wave.
For $\lambda\gtsim 100\micron$, 
the Bruggeman or MG$_2$ EMTs both result in an order of magnitude
greater
suppression of absorption than does our ``standard'' 
Maxwell Garnett approach with $\epsilon_\perp$ inclusions.

The submicron grains in the ISM have typical sizes
$\sim 0.1\micron$.  If the microcrystals are more-or-less
randomly-oriented,
with similar dimensions parallel and perpendicular to $\bchat$,
then we might expect $\sim 1/3$ of the surface area of the grain
to consist of ``basal plane''.  These microcrystals will not
be shielded from incident electric fields that are more-or-less
normal to the grain surface, and thus these ``exposed basal-plane'' 
microcrystals would contribute to absorption. 
If the microcrystal dimensions are
$\sim0.01\micron$, then a significant fraction $\sim0.3$ of the grain volume
is contributed by microcrystals at the grain surface.
We might then expect the far-infrared absorption cross section of
such a particle to be $\sim30\%$ as large as estimated by the 
\onethirdtwothird\ approximation for a single-crystal grain -- 
this happens to be
about what our ``standard'' Maxwell Garnett 
approximation (MG$_1$) gives (see Figure \ref{fig:EMT}).
Further note that there may be a tendency for microcrystals near the
surface to
preferentially orient 
with their basal planes $\sim$parallel to the grain surface,
as in the onion-like presolar graphite grains found in primitive meteorites
\citep[see, e.g.,][]{Bernatowicz+Cowsik+Gibbons+etal_1996}, in which case
the 
$\epsilon_\parallel$ component would be able to
contribute to the absorption without intervening
``shielding'' by $\epsilon_\perp$.

Neither the Bruggeman nor the Maxwell Garnett approaches are theoretically
compelling.  Based on the above discussion,
we tentatively
recommend the MG$_1$ estimate (\ref{eq:MG1}) for the effective
dielectric function of turbostratic graphite grains, but it must
be recognized that this is uncertain.
Theoretical progress on this question appears to require
ambitious calculations using the DDA to solve for the electromagnetic
field within turbostratic graphite structures.
The feasibility of such calculations is uncertain because 
$\epsilon_\parallel$ and especially $\epsilon_\perp$ become numerically large
at wavelengths
$\lambda \gtsim 10\micron$,
making DDA calculations especially challenging.
Alternatively, lab measurements 
of turbostratic graphite grain opacities
at wavelengths $\gtsim30\micron$ 
could determine whether they
agree better with the the Bruggeman and MG$_2$ predictions or the
order-of-magnitude larger opacities predicted by the MG$_1$ EMT.

\begin{figure}[ht]
\begin{center}
\includegraphics[angle=270,width=12.0cm,
                 clip=true,trim=0.5cm 0.5cm 0.5cm 0.5cm]
{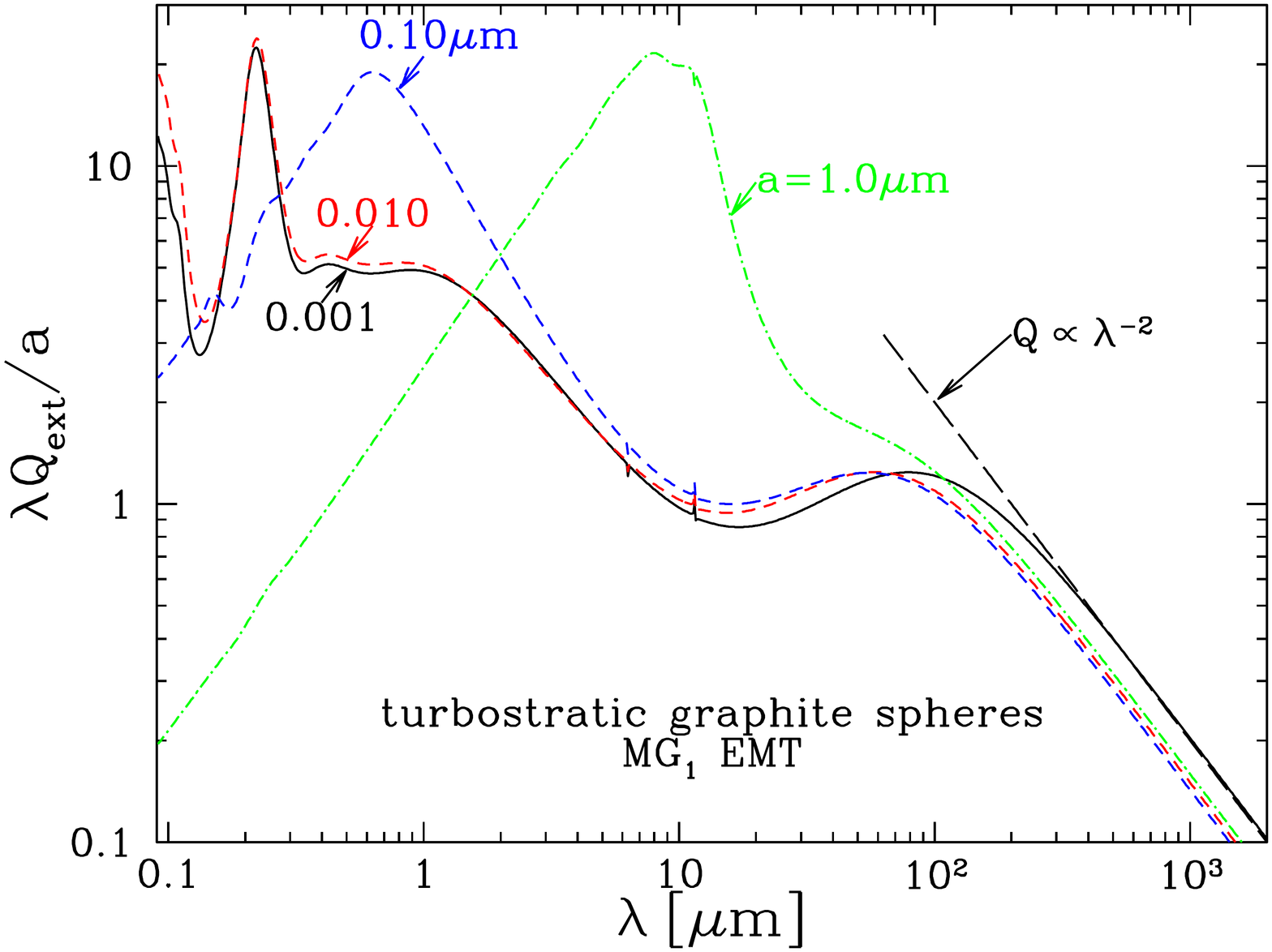}
\caption{\label{fig:qemt_1.000}\footnotesize
         Extinction cross sections for
         turbostratic graphite spheres with selected radii
         $a=0.001\micron$, $0.01\micron$, $0.1\micron$, and $1\micron$.
         Tabulated results for these and other sizes are
         available at \website\ and \websiteb\ .}
\end{center}
\end{figure}

\section{Cross Sections for Turbostratic Graphite Spheres and Spheroids
         \label{sec:emt cross sections}
         }

Adopting the Maxwell Garnett EMT (\ref{eq:MG1}) to estimate an
effective dielectric function $\epsilon_{\rm MG_1}$, we have calculated cross
sections for absorption and scattering by turbostratic graphite
spheres with radii $a$ from $0.001\micron$ to $10\micron$, and
wavelengths from $1\cm$ to $10\Angstrom$.  
Figure \ref{fig:qemt_1.000} shows 
$\lambda Q_\ext/a$ for selected radii.\footnote{%
   Tabulated cross sections are available from \website\ and \websiteb\ .}

\begin{figure}[t]
\begin{center}
\includegraphics[angle=0,width=8.0cm,
                 clip=true,trim=0.5cm 5.0cm 0.5cm 2.5cm]
{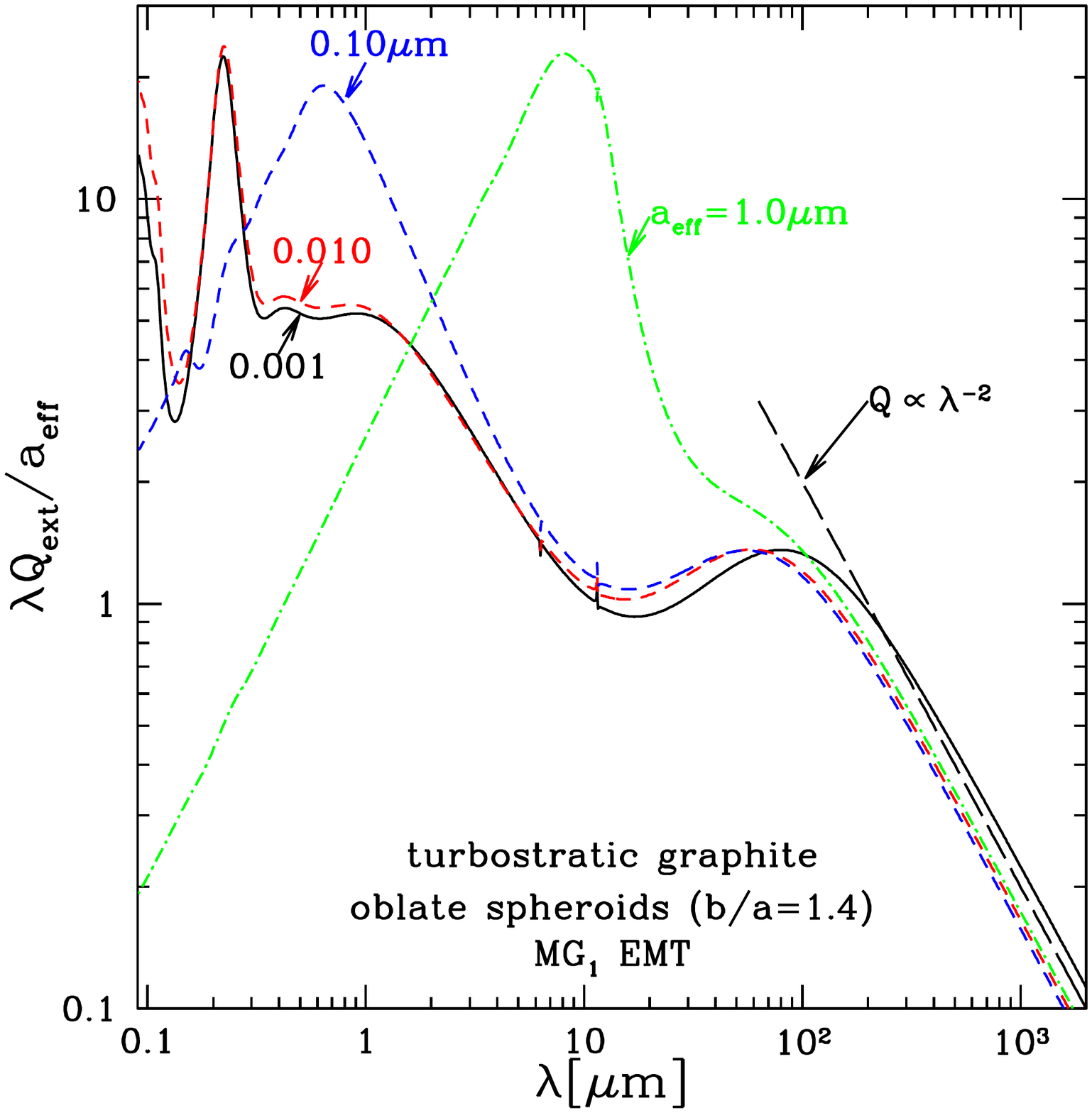}
\includegraphics[angle=0,width=8.0cm,
                 clip=true,trim=0.5cm 5.0cm 0.5cm 2.5cm]
{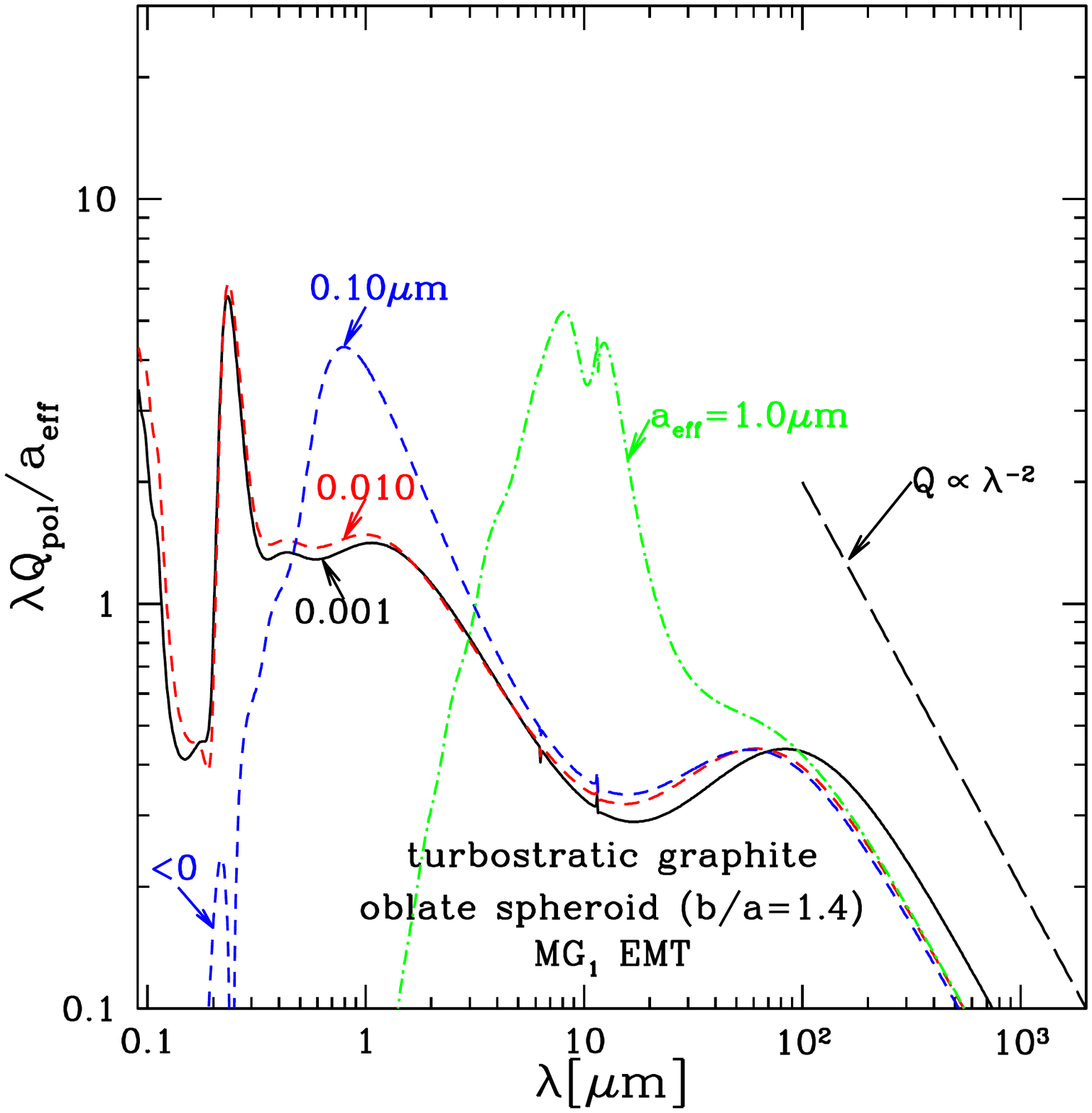}
\caption{\label{fig:pol_1.400}\footnotesize
         Extinction and polarization cross sections for
         oblate spheroids with axial ratio $b/a=1.4$.
         For $a=0.1\micron$, $Q_\pol<0$ for $\lambda<0.25\micron$;
           $\lambda|Q_\pol|/\aeff$ is shown.
         Tabulated cross sections are available at \website\ and
         \websiteb\ .}
\end{center}
\end{figure}

\begin{figure}[t]
\begin{center}
\includegraphics[angle=0,width=8.0cm,
                 clip=true,trim=0.5cm 5.0cm 0.5cm 2.5cm]
{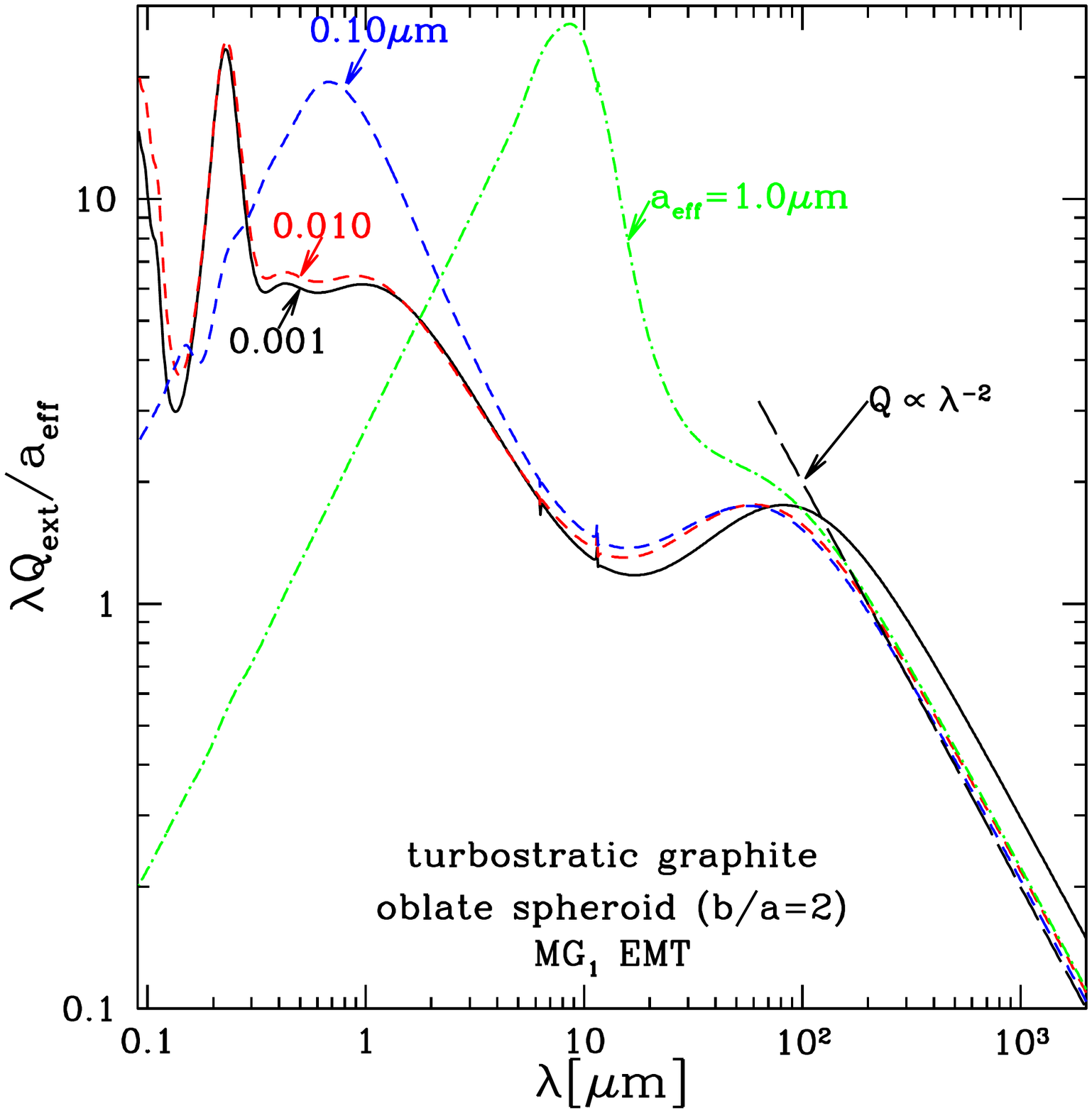}
\includegraphics[angle=0,width=8.0cm,
                 clip=true,trim=0.5cm 5.0cm 0.5cm 2.5cm]
{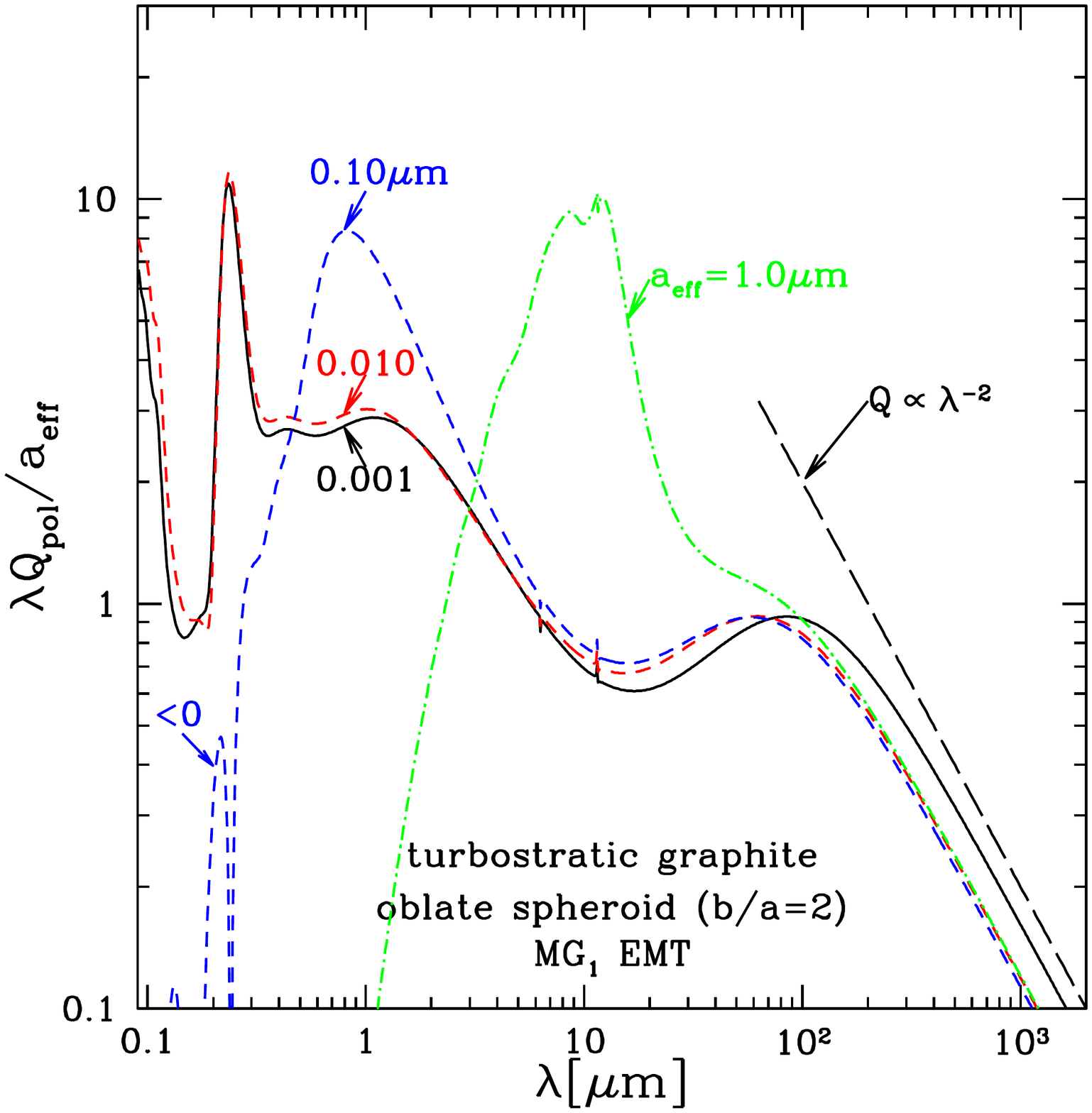}
\caption{\label{fig:pol_2.000}\footnotesize
         Extinction and polarization cross sections for
         oblate spheroids with axial ratio $b/a=2$.
         For $a=0.1\micron$, $Q_\pol<0$ for $\lambda<0.25\micron$;
           $\lambda|Q_\pol|/\aeff$ is shown.
         Tabulated cross sections are available at \website\ and
         \websiteb\ .}
\end{center}
\end{figure}

We have also calculated cross sections for spheroids, with
sizes $\aeff$ ranging from $3.16\Angstrom$ to $5.01\micron$,
various axial ratios $b/a$, and wavelengths $\lambda$ from $1\cm$ to
$0.0912\micron$,
using small-spheroid approximations for $\aeff/\lambda \ll 1$
and the spheroid code developed by \citet{Voshchinnikov+Farafonov_1993} for 
finite $\aeff/\lambda$.
Figures \ref{fig:pol_1.400} and \ref{fig:pol_2.000} show results for
oblate spheroids with axial ratios $b/a=1.4$ and $2$.
The extinction cross section for randomly-oriented dust is estimated
to be
\beq
Q_\ext = \frac{1}{3}
\left[
Q_\ext(\bkhat\!\parallel\!\bahat,\behat\!\perp\!\bahat)+
Q_\ext(\bkhat\!\perp\!\bahat,\behat\!\parallel\!\bahat)+
Q_\ext(\bkhat\!\perp\!\bahat,\behat\!\perp\!\bahat)
\right]
~~~.
\eeq
For perfectly-aligned spheroids, spinning around $\bahat$, the
polarization cross section is defined to be
\beq
Q_\pol \equiv \frac{1}{2}
\left[
Q_\ext(\bkhat\!\perp\!\bahat,\behat\!\perp\!\bahat)-
Q_\ext(\bkhat\!\perp\!\bahat,\behat\!\parallel\!\bahat)
\right]
~~~.
\eeq

\begin{figure}[ht]
\begin{center}
\includegraphics[angle=0,width=8.0cm,
                 clip=true,trim=0.5cm 5.0cm 0.5cm 2.5cm]
{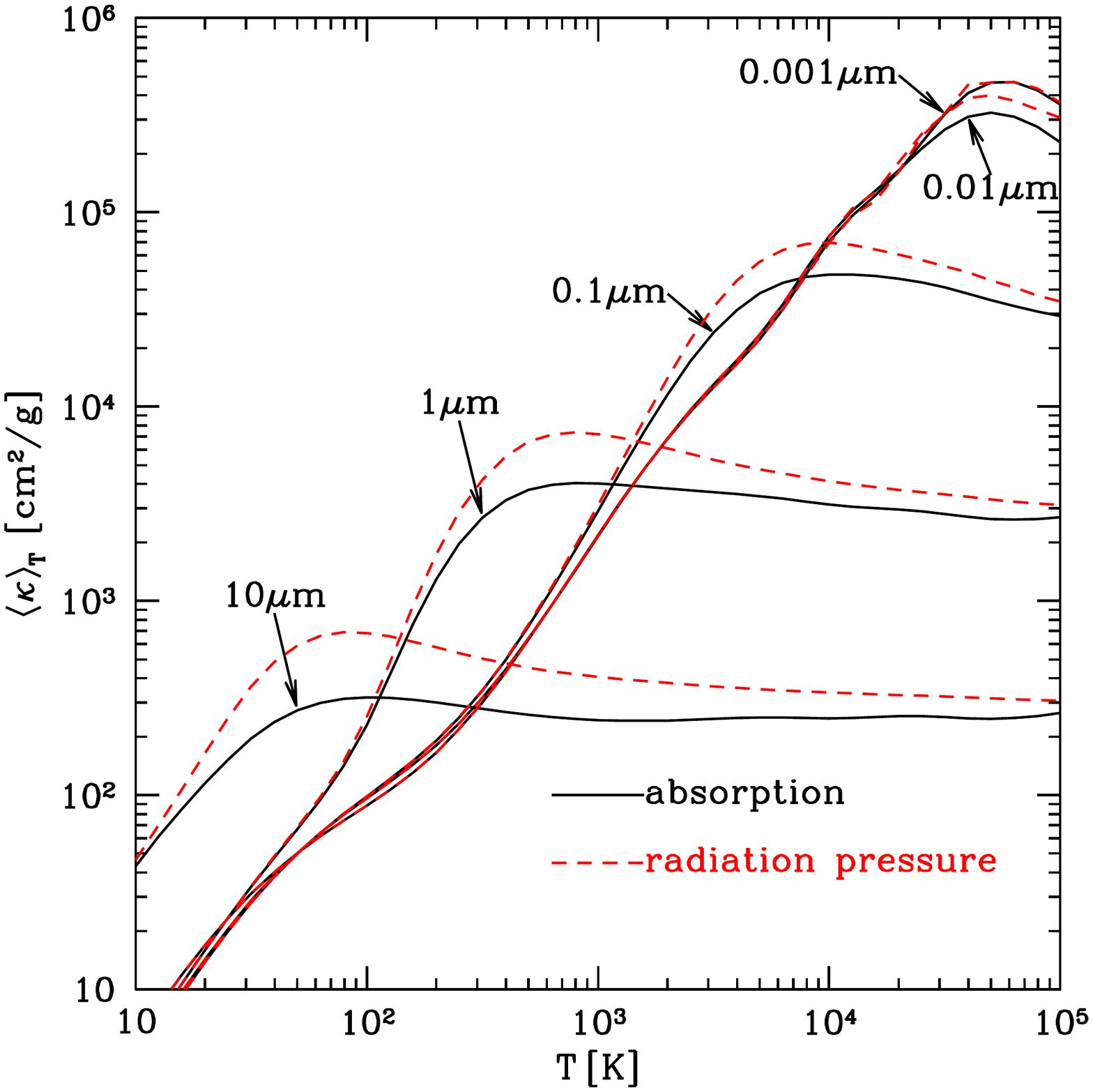}
\caption{\label{fig:planck}\footnotesize
         Planck-averaged opacities for absorption and radiation pressure
         for turbostratic graphite spheres, using the Maxwell Garnett
         EMT.
         Planck-averaged opacities for these and other sizes are
         available at \website\ and \websiteb\ .}
\end{center}
\end{figure}

We have also calculated Planck-averaged cross sections for
selected grain sizes, and temperatures from $10\K$ to $10^5\K$.
Figure \ref{fig:planck} shows Planck-averaged opacities for
absorption and for radiation pressure: 
\beqa
\langle\kappa_\abs\rangle_T &\equiv& \frac{3}{4\rho a}
\frac{\int Q_\abs(\lambda) B_\lambda(T)d\lambda}{\int B_\lambda(T) d\lambda} 
\\
\langle\kappa_{\rm pr}\rangle_T &\equiv& \frac{3}{4\rho a}
\frac{\int \left[Q_\abs(\lambda)+(1-\langle\cos\theta\rangle)Q_\sca)
\right]B_\lambda(T)d\lambda}{\int B_\lambda(T) d\lambda} 
~~~.
\eeqa

\section{\label{sec:lattice resonance}
            Lattice Resonances}

\begin{figure}[ht]
\begin{center}
\includegraphics[angle=270,width=12.0cm,
                 clip=true,trim=0.5cm 0.5cm 0.5cm 0.5cm]
{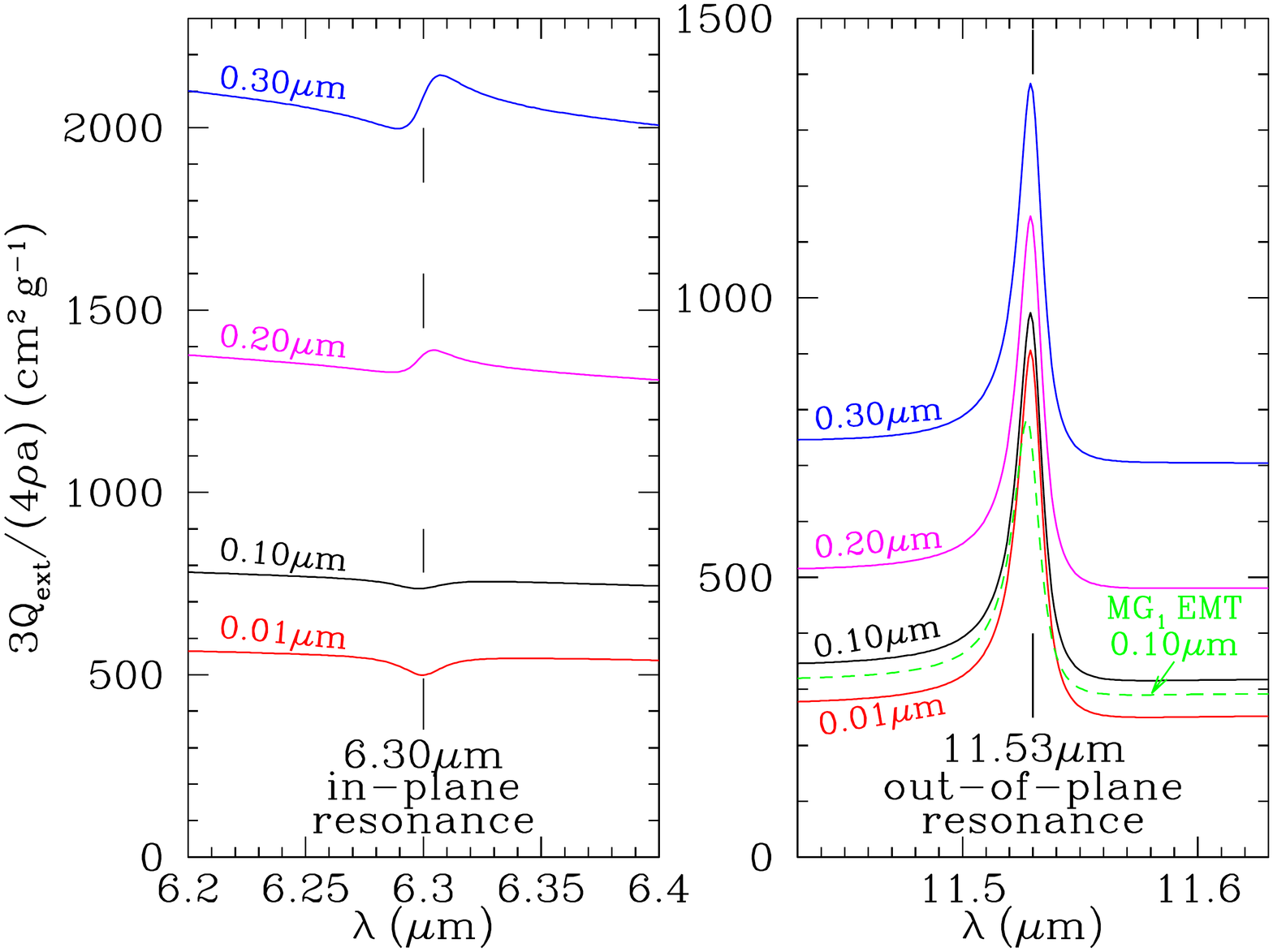}
\caption{\footnotesize\label{fig:reson}
         Extinction opacity for randomly-oriented graphite spheres
         in the neighborhood of the lattice resonances in graphite.
         The $6.30\micron$ in-plane resonance 
         is probably too weak to be observed.
         The $11.5\micron$ out-of-plane resonance is an extinction
         peak, with $\Delta\kappa\approx 640\cm^2\gm^{-1}$ for
         single-crystal spheres; turbostratic graphite grains modeled
         with the Maxwell Garnett EMT (green dashed curve)
         have $\Delta\kappa\approx 470\cm^2\gm^{-1}$ and
         ${\rm FWHM}\approx 0.014\micron$
         }
\end{center}
\end{figure}

\citet{Draine_1984} noted that if graphite is present in the ISM,
the two optically-active lattice resonances, at 6.30 and 
$11.5\micron$,
could conceivably be detected in the interstellar extinction.
Here we reexamine this using the dielectric function developed in this
paper.
An expanded view of the extinction opacity
in the neighborhood of the
resonances is shown in Figure \ref{fig:reson}.  
The $6.30\micron$ in-plane feature
is quite weak, and changes from being an
extinction minimum for small grains, to a complicated profile
for $a\gtsim 0.1\micron$ grains where scattering is no longer negligible.
After averaging over a size distribution, the prospects for detecting
the $6.30\micron$ feature do not seem favorable.
Observability of the feature is further complicated by the presence
of an interstellar absorption feature at $6.2\micron$ that
is likely due to similar C-C stretching modes in aromatic hydrocarbons
\citep{Schutte+vanderHucht+Whittet+etal_1998,Chiar+Tielens+Adamson+Ricca_2013}.

The $11.53\micron$ out-of-plane feature, on the other hand, is 
more prominent as an extinction excess, and the profile is
relatively stable across the range of grain sizes expected in the ISM.

The profiles in Figure \ref{fig:reson} were computed using resonance
parameters measured for HOPG at room temperature.
For randomly-oriented
single-crystal grains, the out-of-plane feature peaks at $11.53\micron$, 
with profile strength
$\Delta\kappa \approx 640\cm^2\gm^{-1}$
and ${\rm FWHM}=0.013\micron$.
The feature is relatively narrow, with $\lambda/{\rm FWHM}\approx 870$.
If the grains are turbostratic graphite modeled using the Maxwell Garnett
effective dielectric function (Eq.\ \ref{eq:MG1}), the profile is somewhat
weaker,
with $\Delta\kappa\approx 470\cm^2\gm^{-1}$ (see Figure \ref{fig:reson}).

Lab studies and ab-initio modeling of the temperature dependence
of the in-plane resonance
\citep{Giura+Bonini+Creff+etal_2012} predict a small frequency shift
$\delta\omega/\omega\approx 0.0025$ as $T$ is reduced from $T=293\K$
to $T\approx20\K$ (appropriate for interstellar grains).
There do not appear to be studies of the $T$-dependence of the
$11.53\micron$
out-of-plane mode, but if the fractional frequency shift is similar,
the $11.53\micron$ peak might shift to $\sim$$11.50\micron$.

If a fraction $\fCgra$ of interstellar
carbon is in graphite, then we expect the resonance feature to have
a maximum optical depth
\beq
\Delta\tau_{11.5}
= 8.3 \times 10^{-25} \left(\frac{\fCgra}{0.3}\right) \cm^2\NH
\eeq
By contrast, the broad silicate feature has 
$\Delta\tau_{9.7}=2.7\times10^{-23}\cm^2\NH$
\citep{Roche+Aitken_1984}.
Thus, the peak optical depth of the graphite feature would be
\beq
\Delta\tau_{11.5}=0.031 \left(\frac{\fCgra}{0.3}\right) \Delta\tau_{9.7}
~~~;
\eeq
the feature is not expected to be very strong.
Spectra taken with the Spitzer IRS-LRS with $R\approx 100$ would
have diluted this by a factor of $\sim10$, and the feature would not
have been detectable even in high S/N IRS-LRS spectra.

The MIRI spectrograph
on JWST, with $R\approx 2500$ near $11.5\micron$, will be well-suited for
study of this feature.
If the resonance parameters for HOPG apply to interestellar graphite,
MIRI may be able to detect interstellar graphite, or obtain
useful upper limits on its abundance, from high S/N spectra of
stars seen through extinction with $\Delta\tau_{9.7}\gtsim 1$.

\section{\label{sec:X-ray}
         X-Ray Absorption and Scattering by Graphite Grains}

\subsection{Randomly-Oriented Grains}
\begin{figure}[ht]
\begin{center}
\includegraphics[angle=0,width=8.0cm,
                 clip=true,trim=0.5cm 0.5cm 0.5cm 0.5cm]
{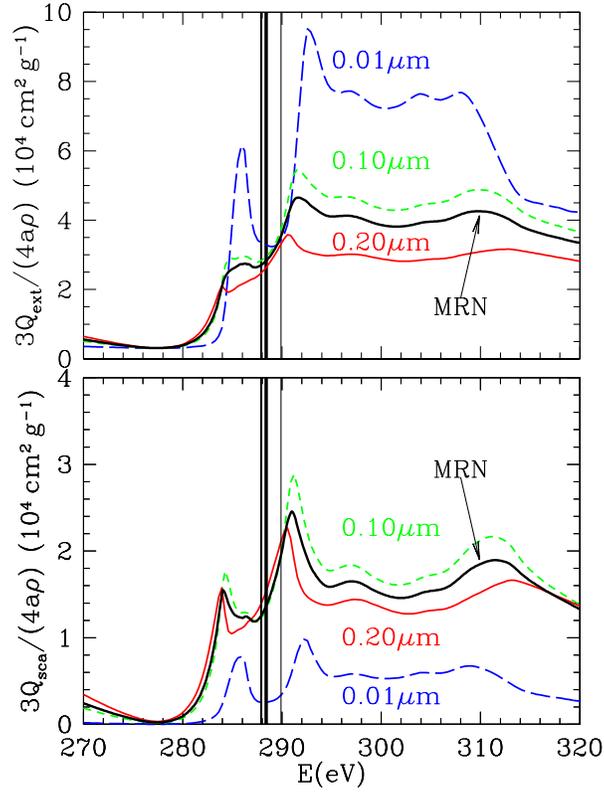}
\caption{\label{fig:kedge} \footnotesize
         Extinction and scattering opacities for
         randomly-oriented
         graphite spheres.
         Broken curves: selected sizes.
         Solid curves: MRN distribution, $dn/da \propto a^{-3.5}$,
         for $0.005\micron < a < 0.25\micron$.
         Vertical lines: positions of strong 
         $1s^22s^22p\rightarrow1s2s^22p^2$ absorption
         lines of \ion{C}{2} (see text).
         }
\end{center}
\end{figure}

X-ray absorption and scattering by carbonaceous dust
near the C K edge can be important
on moderately-reddened sightlines.
Figure \ref{fig:kedge}
shows the extinction and scattering opacities
for randomly-oriented
graphite spheres with three selected radii,
and also 
averaged over two size distributions: (1) the ``MRN'' size distribution,
$dn/da \propto a^{-3.5}$ for $0.005\micron < a < 0.25\micron$
from \citet{Mathis+Rumpl+Nordsieck_1977},
and (2) the size distribution for carbonaceous grains from
\citet{Weingartner+Draine_2001a}.

The cross sections for X-ray scattering and absorption were
calculated
using the \onethirdtwothird\ approximation and
the Mie theory code of \citet{Wiscombe_1980}.
Because the dielectric function is close to unity, we expect
the \onethirdtwothird\ approximation to be accurate at X-ray energies,
even in the present application, where the grain radii $a\approx0.1\micron$
are large compared to the wavelength $\lambda\approx 0.004\micron$.

If 
$A_V/{\rm mag}=5.34 (\NH/10^{22}\cm^{-2})$,
then the graphite grains contribute a mass surface
density
$\Sigma_{\rm C,gra} = 3.73\times10^{-6}(\fCgra/0.3)(A_V/{\rm mag}) \gm\cm^{-2}$.

With $\kappa_{\rm ext}(292\eV)\approx 4\times10^4\cm^2\gm^{-1}$
for the MRN size distribution,
this gives an optical depth 
\beq
\tau_{\rm ext}(292\eV)\approx 0.15\left(\frac{\fCgra}{0.3}\right)A_V
~~~,
\eeq
thus X-ray spectroscopy with moderate signal-to-noise and energy
resolution of a few eV would be able to detect the broad
extinction feature due to the K shell on sightlines with
$A_V\gtsim 1$.
The extinction should peak at $\sim292\eV$; with energy resolution
of $\sim2\eV$, this peak can be separated from
the three \ion{C}{2}\,$1s\!\rightarrow\!2p$ absorption lines (see below).
Unfortunately, the graphite
absorption features which are sharp for small ($\sim 0.01\micron$)
grains are suppressed by internal absorption in the larger grains that
dominate the mass in the MRN distribution,
and the resulting spectroscopic signature is not very pronounced.

For the MRN distribution, about 40\% of the extinction near the K edge
comes from
scattering.  Spectroscopy of the scattering halo would show
an intensity maximum at $\sim291\eV$, and a secondary maximum
at $\sim284\eV$.

In the diffuse ISM, most of the gas-phase carbon
is singly-ionized.
Ground state \ion{C}{2} has strong 
$1s^22s^22p \,{\rm ^2P}^o\rightarrow1s2s^22p^2 
({\rm ^2D}, {\rm ^2P}, {\rm ^2S})$ absorption lines at
$E=287.9\eV$, $288.4\eV$, and $289.9\eV$
\citep{Jannitti+Gaye+Mazzoni+etal_1993,
       Schlachter+SantAnna+Covington+etal_2004},
with oscillator strengths $f=0.102$, $0.193$, and $0.0197$
\citep{Wang+Zhou_2007}.
The resulting $1s2s^22p^2$ excited states (${\rm ^2D}, {\rm ^2P}, {\rm ^2S}$) 
generally deexcite via the
Auger effect, resulting in relatively large intrinsic linewidth
$\Delta E\approx 0.1\eV$.
The lines, shown in Figure \ref{fig:kedge}, will appear as strong
absorption features near $288\eV$, together with the broader features
due to dust.

\begin{figure}[t]
\begin{center}
\includegraphics[angle=0,width=8.0cm,
                 clip=true,trim=0.5cm 5.0cm 0.5cm 2.5cm]
{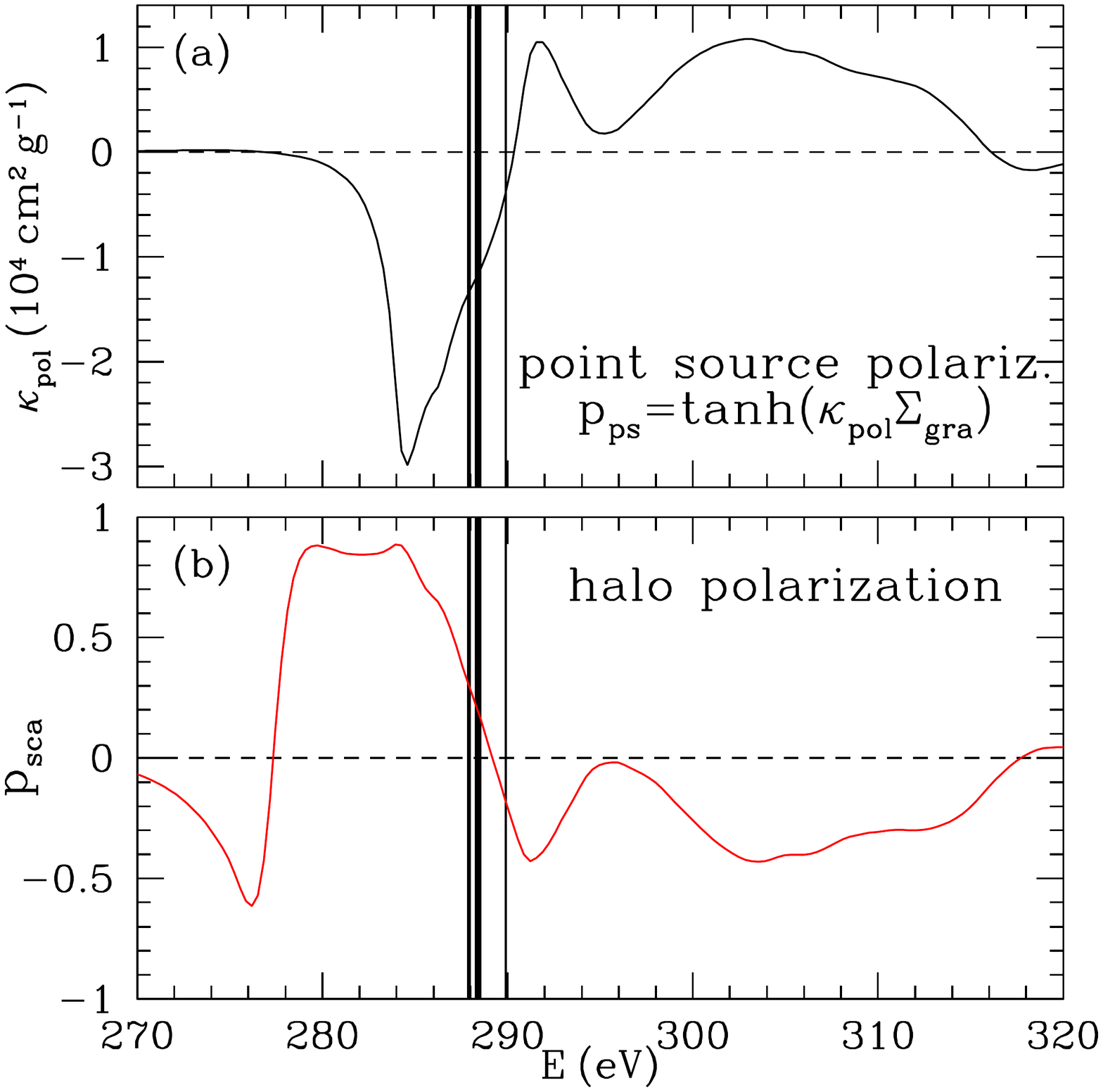}
\caption{\label{fig:kpol} \footnotesize
         Polarization opacity $\kappa_{\rm pol}$
         for single-crystal graphite grains
         with $\aeff=0.1\micron$ and $\bchat\perp\bkhat$.
         $p>0$ for $\bE\parallel\bchat$.
         The scattered halo polarization $p_{\rm sca}>0$
         when $\bE_{\rm sca}\parallel\bchat$.
         Vertical lines: positions of strong K$\alpha$
         $1s^22s^22p\rightarrow1s2s^22p^2$ absorption
         lines of \ion{C}{2} (see text).}
\end{center}
\end{figure}

\subsection{Aligned Graphite Grains: Polarized Extinction and Scattering}

Suppose that the Galactic magnetic field $\bB_0$ is perpendicular
to the line-of-sight to an X-ray source, and that a fraction 
$f_{\rm gra,algn}$
of all of the carbon on the sightline is present in graphite crystals
with $\bchat\parallel\bB_0$; the remaining $1-f_{\rm gra,algn}$
of the C atoms are either not in graphite, or are in randomly-oriented
graphite crystals.

If the total carbon abundance is C/H=295ppm,
and $\NH=1.87\times10^{21}\cm^{-2}A_V/{\rm mag}$, then
radiation reaching us from a point source seen through dust
with extinction $A_V$ will acquire a polarization
\beq
P_{\rm ps} = \tanh\left[\kappa_{\rm pol}\times
1.12\times10^{-5}\gm\cm^{-2}f_{\rm gra,algn}
\left(\frac{A_V}{\rm mag}\right)\right]
~~~.
\eeq
With $\kappa_{\rm pol}(285\eV)\approx -3\times10^4\cm^2\gm^{-1}$
(see Fig.\ \ref{fig:kpol}),
\beq
P_{\rm ps}(285\eV) = 
\tanh(-0.35 f_{\rm gra,algn}A_V)\approx-0.35f_{\rm gra,aligned}(A_V/{\rm mag})
~~~.
\eeq

Thus if, say, 10\% of the carbon were in aligned graphite crystals
with $\bchat\perp$ line-of-sight
on a sightline with $A_V=1$,
the polarization of the point source
would be 3.5\% at $285\eV$.
While the polarization would increase for increased $A_V$, the
overall attenuation by the ISM makes it difficult to carry out
observations near the C K edge on sightlines with $A_V\gtsim 2$. 

For oriented graphite grains, the scattered X-rays
would show energy-dependent polarization, as in 
Fig.\  \ref{fig:kpol}.
The X-ray scattered halo 
could achieve polarizations as large as 85\% at $284\eV$, 
although this would
be reduced when the effects of polarized absorption, and
unpolarized scattering by nonaligned
or nongraphitic grains, are included.
Because the polarization is strongly energy-dependent (see
Figure \ref{fig:kpol}) an instrument intended to measure it should have
$\ltsim10\eV$ energy resolution.

At this time there are no instruments, existing or planned,
that could measure such X-ray polarization.  
Proposed
X-ray polarimetry missions
(GEMS and IXPE) are intended to operate only above 2 keV,
and have minimal energy resolution.

\section{\label{sec:discussion}
         Discussion}

The dielectric function of graphite continues
to be uncertain, which is surprising for such a well-defined
and fundamental material.  
In particular, there are striking differences in the reported
optical constants for $\bE\parallel\bchat$ in the 1--5$\eV$ range for, as
seen in Figure \ref{fig:epsilon_para}, and there do not appear
to be any published measurements between 0.3 and 1$\eV$.
The synthetic dielectric functions obtained here represent our best
effort to reconcile the published experimental results.
We hope that there will be renewed
efforts to accurately 
measure both $\epsilon_\perp$ and $\epsilon_\parallel$ at
wavelengths from the infrared to the ultraviolet.

Of particular interest would be study of the
11.5$\micron$ lattice resonance at temperatures $T\approx20\K$ appropriate
to interstellar grains, to determine the precise wavelength where this
resonance should be seen if the interstellar grain population contains
a significant amount of graphite.  It would also be desirable to study
this absorption resonance in disordered forms of carbon that contain
microcrystallites of graphite, so that detection of, or upper limits
on, the presence of an $11.5\micron$ feature in spectra obtained with
MIRI can be used to determine the amount of graphitic carbon in the
ISM.

The optical properties of turbostratic graphite at wavelengths 
$\lambda\gtsim10\micron$ remain very uncertain.
Different effective medium theories make very different predictions
for the effective dielectric function that is intended to characterize
turbostratic graphite.
Theoretical progress requires obtaining accurate solutions to
Maxwell's equations for particles consisting of turbostratic
graphite material.  The discrete dipole approximation is one possible
numerical technique, but at infrared wavelengths
the very large dielectic function 
for $\bE\perp\bchat$ makes DDA calculations numerically
challenging.
Alternatively, it may be possible to carry out laboratory measurements
of absorption by turbostratic graphite particles, to compare with
the predictions of different effective medium theories.

\section{\label{sec:summary}
         Summary}

The principal results of this paper are as follows:
\begin{enumerate}
\item 
A dielectric tensor for graphite is presented,
extending from zero frequency to X-ray energies, and based on
laboratory data.  Except for the absorption
by the K shell electrons at $E>280\eV$, $\epsilon_\perp$ and
$\epsilon_\parallel$ are
given by easily-evaluated analytic expressions
(Eq.\ \ref{eq:epsilon} and Tables \ref{tab:parameters for Eperpc},
\ref{tab:parameters for Eparac}).
 
\item
Techniques for calculating absorption and scattering cross sections
are discussed.
For single-crystal graphite grains, 
the simple \onethirdtwothird\ approximation is exact for $a/\lambda \ll 1$,
and is shown to be moderately
accurate even when $a/\lambda$ is not small.
For $0.1\ltsim\lambda/\micron<0.5$, the \onethirdtwothird\ approximation
gives cross sections that are accurate to within $\pm5\%$, at least
for spheres (see Figure \ref{fig:dda_sphere_test}.
At infrared wavelengths, where $\epsilon_\perp$ becomes large, 
the \onethirdtwothird\ approximation tends to
overestimate $Q_\ext$ when $a/\lambda \ltsim 0.1$, and to underestimate
$Q_\ext$ when $0.1< a/\lambda \ltsim 0.2$.
For $\lambda=10\micron$, the \onethirdtwothird\ approximation may
overestimate $Q_\ext$ by as much as 40\%.
If errors of tens of percent are tolerable, the 
\onethirdtwothird\ approximation can be used
when more exact DDA calculations are unavailable.
\item If a significant fraction of interstellar carbon is in the form
of graphite, 
the $11.5\micron$ lattice resonance may be detectable with
$R\gtsim 10^3$ spectroscopy (see Fig.\ \ref{fig:reson}).
The MIRI spectrograph
on JWST will be able to detect or obtain useful upper limits on
the 11.5$\micron$ feature.
\item For grains consisting of turbostratic graphite
(randomly-oriented graphite microcrystallites),
either the Maxwell Garnett or Bruggeman theories can
be used to obtain an effective dielectric function for use in
scattering calculations at wavelengths $\lambda\ltsim 0.5\micron$.
However, at longer wavelengths, the Maxwell Garnett and Bruggeman estimates
diverge.  We suggest that 
one of the Maxwell Garnett estimates -- 
MG$_1$ given by Eq.\ (\ref{eq:MG1}) --
may be the best choice for turbostratic graphite particles.
However, the applicability of effective medium theory to
turbostratic graphite at $\lambda\gtsim10\micron$ remains uncertain,
and additional theoretical and/or experimental work is required.
\item 
The carbon K shell absorption in graphite is anisotropic.  If 
interstellar carbon were substantially in small, aligned graphite grains,
the K edge absorption would result in significant polarization of the
transmitted X-rays between $280$ and $310\eV$.
The scattered X-ray halo produced by aligned grains
will be oppositely polarized (see Fig.\ \ref{fig:kpol}).
Currently planned X-ray telescopes lack polarimetric capabilities
near the carbon K edge, but future X-ray observatories might
be able to detect or constrain such polarization.
\end{enumerate}

\acknowledgments
I am grateful to
N.~V.\ Voshchinnikov for generously making available the program
HOM6\_4N4 for calculating absorption and scattering by isotropic spheroids;
to R.~A.\ Rosenberg
for helpful communications, and for permission to reproduce Figure 1
from \citet{Rosenberg+Love+Rehn_1986},
and
to B.\ Hensley 
and D.\ Gutkowicz-Krusin
for helpful remarks.
I thank the anonymous referee for very helpful comments and suggestions.
This work was supported in part by NSF grant AST-1408723.

\bibliography{/u/draine/work/bib/btdrefs}
\end{document}